\documentclass[submission, Phys]{SciPost}

\usepackage[utf8]{inputenc} 
\usepackage[T1]{fontenc} 	
\usepackage[english]{babel} 

\usepackage[bitstream-charter]{mathdesign}
\usepackage{geometry} 		
\usepackage{amsmath} 		
\usepackage{mathtools} 		
\usepackage{float} 			
\usepackage{graphicx} 		
\usepackage{tabularx} 		
\usepackage{booktabs} 		
\usepackage{color, xcolor} 	
\usepackage{pdfpages} 		
\usepackage{extarrows} 		
\usepackage{multirow} 		
\usepackage{multicol} 		
\usepackage{enumitem} 		
\usepackage{xspace} 		
\usepackage{stackrel} 		
\usepackage{tikz} 			
\usepackage{braket} 		
\usepackage{bm} 			
\usepackage{tensor} 		
\usepackage{slashed} 		
\usepackage{siunitx} 		
\usepackage{lastpage} 		
\usepackage{cite} 			
\usepackage[normalem]{ulem} 
\usepackage{fontawesome} 	
\usepackage{tocloft} 		
\usepackage{titlesec} 		
\usepackage{doi} 			
\usepackage{hyperref} 		
\usepackage[most]{tcolorbox} 					
\usepackage[nameinlink, capitalize]{cleveref} 	
\usepackage[nottoc, notlot, notlof]{tocbibind} 	
\usepackage[ruled, vlined]{algorithm2e} 		
\usepackage{makecell}
\usepackage[makeroom]{cancel}
\usepackage{feynmf}

\makeatletter
\def\BState{\State\hskip-\ALG@thistlm}
\makeatother

\makeatletter
\@ifundefined{pdfoutput}{}{\DeclareGraphicsRule{*}{mps}{*}{}}
\makeatother

\makeatletter
\DeclareRobustCommand*{\bfseries}{%
   \not@math@alphabet\bfseries\mathbf
   \fontseries\bfdefault\selectfont
   \boldmath
}
\makeatother

\hypersetup{
	pdftitle={T3uncertainty},
	pdfauthor={Elmer et al.},
	colorlinks=true, 			
	linkcolor={red!50!black}, 	
	citecolor={blue!50!black}, 	
	urlcolor={blue!80!black} 	
} 

\DeclareSymbolFont{usualmathcal}{OMS}{cmsy}{m}{n}
\DeclareSymbolFontAlphabet{\mathcal}{usualmathcal}

\SetArgSty{textnormal}
\SetKwComment{Comment}{{\small\#}~}{}
\SetCommentSty{mycommfont}

\setitemize{itemsep=0pt, parsep=0pt} 				
\setenumerate{itemsep=0pt, parsep=0pt} 				
\setitemize{itemsep=2pt,topsep=2pt,parsep=0pt,partopsep=0pt,leftmargin=*}
\setenumerate{itemsep=0pt,topsep=2pt,parsep=0pt,partopsep=0pt,labelindent=3pt,leftmargin=*}
\usepackage{amsmath}
 
\usepackage{amsthm} 		
\theoremstyle{definition}

\newlist{todolist}{itemize}{2}
\setlist[todolist]{label=$\square$}
\usepackage{pifont}
\newcolumntype{C}[1]{>{\centering\arraybackslash}p{#1}}
\newcolumntype{P}[1]{>{\centering\arraybackslash}p{#1}}

\newcommand{\be}{\begin{equation}}
	\newcommand{\ee}{\end{equation}}
\newcommand{\bea}{\begin{eqnarray}}
	\newcommand{\eea}{\end{eqnarray}}
\newcommand{\bi}{\begin{itemize}}
	\newcommand{\ei}{\end{itemize}}
\newcommand{\ben}{\begin{enumerate}}
	\newcommand{\een}{\end{enumerate}}

\newcommand{\eqcomma}{\;\text{,}} 	
\newcommand{\eqperiod}{\;\text{.}} 	

\graphicspath{{./figs/}}

\begin{document}

\begin{center}
{\Large \textbf{
Faithful Uncertainties in ML-Based Inverse Problems:\\ GPs, NNs and NTKs for the Triplet PDF
}}
\end{center}

\begin{center}
Nina Elmer\textsuperscript{1},
Sven Krippendorf\textsuperscript{1,2},
and Maria Ubiali\textsuperscript{1}
\end{center}

\begin{center}
{\bf 1} DAMTP, University of Cambridge, Wilberforce Road, Cambridge, CB3 0WA, United Kingdom\\
{\bf 2} Cavendish Laboratory, University of Cambridge, J J Thomson Avenue, Cambridge, CB3 0US, United Kingdom\\
\end{center}

\begin{center}
\today
\end{center}


\section*{Abstract}
{\bf 
Machine-learning methods are increasingly used to solve inverse problems in physics, where a faithful estimate of the uncertainty on the reconstructed function is as important as its central value. We address this question in the simplest instance of a parton distribution function (PDF) determination: the reconstruction of the non-singlet triplet distribution $T_3(x,Q_0^2)$ from deep-inelastic scattering data. Within a closure-test framework, in which the underlying PDF is known and the pseudo-data carry correlated experimental uncertainties, we compare Gaussian processes with different kernels, neural-network ensembles trained with different loss functions, and empirical neural tangent kernel approximations. We find that methods yielding comparable central reconstructions can produce markedly different uncertainties, and we identify the ingredients that control them: a likelihood that includes the full experimental covariance, the functional prior encoded in the kernel or in the endpoint preprocessing, and, for tangent-kernel approximations, the stage of training at which the kernel is evaluated. A normalised-bias diagnostic shows that calibration depends on the correlations of the predictive covariance, not only on its pointwise width. Our results illustrate why closure tests and covariance-aware diagnostics are essential to validate uncertainties in ML-based approaches to inverse problems.
}

\newpage
\vspace{1pt}
\noindent\rule{\textwidth}{1pt}
\tableofcontents\thispagestyle{fancy}
\noindent\rule{\textwidth}{1pt}
\vspace{1pt}

\clearpage

\section{Introduction}
\label{sec:intro}
Machine-learning (ML) methods are increasingly used to solve inverse problems in physics, in which a continuous 
function has to be reconstructed from a finite set of indirect and noisy measurements. 
In such problems, the uncertainty on the reconstructed function is as important as its central value, since it 
determines how the result propagates to predictions and how reliably it can be used in subsequent analyses. 
In high-energy physics, where large datasets make systematic uncertainties dominant, faithful uncertainty 
quantification (UQ) is vital to realise the 1\% accuracy target of the High-Luminosity LHC~\cite{Bruning:2025pmh}. Similar considerations arise in precision cosmology, where reliable parameter constraints require careful propagation of observational uncertainties and validation against modelling and analysis assumptions (cf.~\cite{Planck:2018vyg,DES:2021wwk}). 
In parallel, the ML community has moved from point estimates towards probabilistic predictions based on Bayesian 
frameworks and ensemble methods~\cite{MLgenUQ1,MLgenUQ2,MLgenU3}, and reliable UQ has become a central requirement 
of ML-driven physics research~\cite{Chen:2022pzc,Haussmann:2026gbi}.

The determination of the parton distribution functions (PDFs) of the proton from experimental data is a prime example 
of such an inverse problem~\cite{DelDebbio:2026abr,Ubiali:2024pyg,Ball:2026dyb}. 
Neural networks (NNs) have been used in PDF determinations for over two decades, starting from the first neural-network 
parametrisation of DIS structure functions~\cite{Forte:2002fg} 
and the subsequent NNPDF analyses, whose explicit goal was a determination of PDFs with faithful uncertainty 
estimation~\cite{DelDebbio:2007ee,Ball:2008by,NNPDF:2021njg}. 
To validate this faithfulness, the NNPDF collaboration has developed closure 
tests~\cite{NNPDF:2014otw,DelDebbio:2021whr,Barontini:2023vmr}, in which pseudo-data are generated 
from a known underlying PDF with realistic experimental uncertainties, so that the fitted PDFs and their 
uncertainties can be compared directly with the truth. 
Closure tests have since become an established tool to assess the statistical properties of PDF uncertainties, 
including those of the Monte Carlo replica method used to propagate them~\cite{Costantini:2024wby,Costantini:2026sel}. 
Since they do not depend on the specific methodology being tested, closure tests also provide a natural framework to 
compare different approaches to the same inverse problem on an equal footing~\cite{Harland-Lang:2024kvt}, 
which is the aim of this work.

Among the methods used for such problems, Gaussian processes (GPs)~\cite{rasmussen2006gaussian} play a 
distinguished role: as Bayesian, non-parametric models, they provide predictive distributions with analytically 
tractable uncertainties and have long served as reference models for UQ. 
GPs have been used to solve inverse problems in various fields of physics, from geophysics~\cite{valentine2020gaussian} 
to lattice QCD~\cite{Karpie:2019eiq,Horak:2021syv,Horak:2023xfb,DelDebbio:2024lwm}, and more recently in the 
determination of PDFs~\cite{Candido:2024hjt,Medrano:2025cmg}. 
In the extraction of spectral densities from lattice correlators, the GP solution of the inverse problem has been 
shown to be related to Backus--Gilbert methods~\cite{DelDebbio:2024lwm}, such as that of Ref.~\cite{Hansen:2019idp}. 
A defining feature of GPs is the choice of kernel, which encodes prior assumptions about smoothness, correlations 
and functional structure. 
While this flexibility is one of the strengths of the framework, it raises the question of how strongly predictions 
and their uncertainties depend on the chosen kernel. 
Recent studies have started to address this question by assessing the sensitivity of the GP posterior 
to ``qualitatively interchangeable'' kernels~\cite{stephenson2022measuring} 
or by systematically evaluating kernel compositions~\cite{lu2025optimal}. 
However, systematic studies of how the kernel choice affects uncertainty estimates in physics inverse problems 
remain limited.

NNs are instead deterministic function approximators, whose uncertainty 
has to be estimated from an ensemble of networks: trained either on Monte Carlo replicas of the data, as in the 
NNPDF approach to PDF determination~\cite{Ball:2008by,NNPDF:2021njg,Costantini:2024wby,Costantini:2026sel}, or 
on the same data with different initialisations, 
as in the estimate of the uncertainty of surrogate models of scattering amplitudes~\cite{Bahl:2024gyt,Bahl:2025xvx}. 
The resulting uncertainty depends on how the information on the data uncertainties enters the training, namely on 
the loss function. 
In PDF fits, the loss is the $\chi^2$ built from the full experimental covariance matrix and the theory covariance 
matrix~\cite{NNPDF:2024dpb}. 
For certain surrogate models, as for example used in Refs.~\cite{Bahl:2024gyt, Bahl:2025xvx}, by contrast, no covariance matrix is available, and the data uncertainty is either ignored, as in a mean-squared-error loss, or learned from the data through a heteroscedastic loss.

NNs and kernel methods are closely related. 
In the infinite-width limit, randomly initialised networks converge to GPs governed by architecture-dependent 
kernels~\cite{Neal:1996nn,Lee:2018DeepGP}, 
while their training dynamics under gradient descent is described by the 
Neural Tangent Kernel (NTK)~\cite{Jacot:2018NTK,Lee_2020}. 
At finite width the empirical NTK evolves during training, and its study has recently been used to 
understand the training of NN-based PDF fits and their uncertainties~\cite{DelDebbio:2024qfh,Chiefa:2025cap}. 

We compare these approaches in the simplest PDF determination problem: the extraction of the non-singlet triplet 
PDF $T_3$ from the deep-inelastic scattering (DIS) structure functions measured by the BCDMS experiment~\cite{BCDMS:1994ala}. 
Here a single function is related to the data through a linear forward map, so that the problem retains the key 
features of a PDF determination 
while allowing for a fully controlled comparison of different methodologies. 
Single-flavour determinations of this kind have historically served as a testbed for new methodologies before their 
extension to global analyses: 
the NNPDF approach was first applied to the non-singlet distribution from DIS data~\cite{DelDebbio:2007ee}, then to 
a complete DIS fit~\cite{Ball:2008by}, 
before evolving into the current global analyses that include data from proton-(anti)proton 
colliders~\cite{Ball:2010de,NNPDF:2021njg,NNPDF:2024nan,NNPDF:2024djq}. 
Following the same strategy, $T_3$ has been used as a case study for GP-based PDF 
determinations~\cite{Candido:2024hjt} and for the study of the 
NTK of NN-based PDF fits~\cite{Chiefa:2025cap}.

Building on these studies, we use the $T_3$ problem in a closure-test setting, in which several levels of 
noise can be added to the pseudo-data~\cite{NNPDF:2014otw,DelDebbio:2021whr,NNPDF:2021njg,Barontini:2023vmr}, 
to study uncertainty quantification both in the region constrained
by the data and in the extrapolation region.
Both regions matter for phenomenology.
At the LHC, PDFs are needed over the whole range from $x\sim10^{-5}$ up to $x\sim0.8$:
the large-$x$ region is crucial for the production of heavy particles, for instance in searches for new 
physics at high invariant masses, while the small-$x$ region provides a test of QCD dynamics and is probed 
by processes at large rapidities~\cite{Amoroso:2022eow,Ubiali:2024pyg}.
Since any given dataset constrains only part of this range, a faithful uncertainty is needed both where the data 
lie and where the PDF is extrapolated.
We compare GPs with different kernels, ensembles of NNs trained with different loss functions on different levels 
of closure-test data, 
and NTK-based approximations evaluated at different stages of training. 
We assess the faithfulness of their uncertainties with a normalised-bias diagnostic that accounts for the 
correlations of the predictive covariance. 
We find that methods yielding comparable central reconstructions can produce markedly different uncertainties, 
and that these are controlled by the 
treatment of the experimental covariance in the likelihood, by the functional prior encoded in the kernel or in 
the endpoint preprocessing, 
and, for NTK approximations, by the stage of training at which the kernel is evaluated. 
Finally, we test the reconstructed $T_3$ in the extrapolation region~\cite{Cruz-Martinez:2021rgy} by 
propagating it to HERA neutral-current reduced cross sections, 
which probe values of $x$ not directly constrained by the BCDMS data.

The paper is structured as follows. 
In Sec.~\ref{sec:t3model} we define the inverse problem, the forward map and the levels of closure-test data. 
In Sec.~\ref{sec:setup} we describe the GP, NN and NTK setups, together with the uncertainty estimators and 
diagnostics used in the analysis. 
In Sec.~\ref{sec:results} we compare the resulting reconstructions and test the faithfulness of their uncertainties, 
first in the data (interpolation) region by means of the normalised bias,
and then in the extrapolation region by propagating them to HERA reduced cross sections, which partly lie at 
values of $x$ not covered by the training data.
We conclude and provide an outlook in Sec.~\ref{sec:summary}.

\section{The terms of the inverse problem: data and forward map}
\label{sec:t3model}
In this section we define the inverse problem studied in this work: the determination of the non-singlet triplet PDF $T_3$ from DIS structure functions 
measured by the BCDMS experiment~\cite{BCDMS:1994ala}. 
Although this is a reduced version of a global PDF fit, involving a single flavour combination rather than the full set of seven or eight independent PDFs 
(depending on whether an intrinsic charm component is fitted~\cite{Ball:2022qks}) and a single observable rather than a global dataset, 
it retains the defining features of the general problem of determining the continuous functions that parametrise the proton structure 
from a finite set of lepton--hadron and hadron--hadron scattering data~\cite{DelDebbio:2026abr}. 

In what follows we discuss the general forward map that connects the PDFs to the observables at lepton-proton and proton-proton colliders (Sec.~\ref{subsec:FKPDF}).  
We then describe the specific forward map connecting the triplet PDF to the DIS structure functions (Sec.~\ref{subsec:FKT3}) and give the relevant definitions that characterise the closure test set-up that we adopt (Sec.~\ref{subsec:closure}). 
The architectures, loss functions and uncertainty estimators are introduced in Sec.~\ref{sec:setup}.

\subsection{The inverse problem associated with PDF determinations}
\label{subsec:FKPDF}
The extraction of PDFs from experimental data is a classic example of an inverse problem, namely, the
reconstruction of a set of functions $f_i(x)$ from a finite set of noisy data points $y_n$, with index $n = 1,\ldots,N_{\rm dat}$ 
where $x$ denotes the fraction of the proton momentum carried by the parton~\cite{DelDebbio:2026abr}.
The forward map associated with a PDF analysis is the operator that maps the PDFs at some fixed input scale $Q_0$ 
onto the vector of theory predictions for the observables entering the PDF analysis. The collection of PDFs is denoted by 
$\mathbf{f} \equiv \{f_i(x, Q_0^2)\}_{i=1}^{N_{\mathrm{flav}}}$ at the initial scale, and the forward map
\begin{equation}
  \mathcal{G} : \mathbf{f} \longmapsto
  \big(\mathcal{G}_1[\mathbf{f}], \ldots, \mathcal{G}_{N_{\mathrm{dat}}}[\mathbf{f}] \big) \in \mathbb{R}^{N_{\mathrm{dat}}}
\label{eq:forward-map-def}
\end{equation}
is defined by the convolution of the perturbatively computable Wilson coefficient $C_{n,i}$ with the DGLAP-evolved PDF. 
For a generic DIS observable indexed by $n$, we can set the renormalisation and factorisation scales to $Q^2_n$, where $Q^2_n$ is the virtuality of the exchanged boson. 
This reads as
\begin{equation}
  \mathcal{G}_n[\mathbf{f}] = \sum_{i=1}^{N_{\rm flav}} C_{n,i}\!\otimes
    f_i(Q_n^2) \equiv \sum_{i=1}^{N_{\rm flav}}  \int_{x_n}^{1} \frac{d\xi}{\xi}\,
    C_{n,i}\!\left(\frac{x_n}{\xi},\, \alpha_s(Q_n^2)\right)
    f_i(\xi,Q_n^2) \eqcomma
\label{eq:forward-dis}
\end{equation}
with $\otimes$ the symbol for the integral convolution that is defined in the above equation, and $f_i(\xi,Q_n^2)$ obtained from $f_j(\xi, Q_0^2)$ by the action of the DGLAP evolution
operators 
{\small
\begin{equation}
\label{eq:dglap}
    f_i(\xi,Q_n^2) = \sum_{j=1}^{N_{\rm flav}}\Gamma_{ij}\left(Q_0^2\to Q_n^2\right) \otimes f_j(\xi,Q_0^2) \equiv \sum_{j=1}^{N_{\rm flav}}\int_{\xi}^{1} \frac{d\zeta}{\zeta}\,\Gamma_{ij}\left(\frac{\xi}{\zeta},Q_0^2\to Q_n^2\right) \, f_j(\zeta,Q_0^2) \eqperiod
\end{equation}
}
For collider observables with two hadrons in
the initial state, the analogous expression is bilinear in the PDFs.
\newline
Evaluating the two integral convolutions, Eq.~\eqref{eq:forward-dis} and Eq.~\eqref{eq:dglap} (or the corresponding expressions for hadronic observables), for every data point $n$ at every step of a global PDF fit is computationally prohibitive. Modern pipelines therefore
pre-compute, once and for all, the combined action of the DGLAP evolution, perturbative matching, and
phase-space integration as a set of Fast-Kernel (FK) tables on fixed interpolation grids in
$(x, Q^2)$~\cite{Ball:2010de,Carli:2010rw,Kluge:2006xs,Carrazza:2020gss,Candido:2022tld,Barontini:2023vmr,Candido:2024rkr}. With the FK-tables in hand, 
the theoretical prediction for any data
point reduces, at fixed $\alpha_s$ and fixed scale choices, to a tensor contraction of the FK
array with the parametrised PDFs evaluated at $Q_0$ on the interpolation grid. For DIS 
observables, with a single hadron in the initial state, the integral convolutions of Eqs.~\eqref{eq:forward-dis}-\eqref{eq:dglap} discretise to the FK approximation $G_n$ of $\mathcal{G}_n$,
\begin{equation}
G_n[\mathbf{f}]
  = \sum_{i,\,\alpha}\,W_{n,\,i\alpha}\,
    f_i(x_\alpha, Q_0^2) \eqcomma
\label{eq:fk-dis}
\end{equation}
where $W$ are the FK-tables, $i$ indexes the parton flavour $i=1,\ldots N_{\rm flav}$ and $\alpha=1,\ldots,N_{\rm grid}$ the interpolation node in~$x$. 
The FK array $W$ accounts for the combined action of amplitude calculation, 
DGLAP evolution and phase-space integration, reducing each forward evaluation to a tensor 
contraction and making the iterative optimisation or posterior sampling required by a
global analysis computationally tractable.

\subsection{The determination of the triplet PDF from BCDMS data} 
\label{subsec:FKT3}
For this study, we will focus on a specific DIS observable, which depends linearly on a single PDF, $T_3(x)$, rather than on $N_{\rm flav}$ functions 
as in a generic PDF analysis.  The non-singlet triplet PDF $T_3$ is a specific PDF flavour combination defined as 
\begin{align}
    T_3 = u^+ - d^+ \eqcomma
\end{align}
where $u^+ = u + \bar{u}$ and $d^+ = d + \bar{d}$. 
Considering DIS on a proton target, the theory prediction for the proton structure function $F_2^p$ can be written, at any perturbative order, as 
\begin{align}
    F_2^{p} = C_g \otimes g + C_\Sigma \otimes \Sigma + C_{T_3} \otimes T_3 + C_{T_8} \otimes T_8 + C_{T_{15}} \otimes T_{15} \eqcomma
\end{align}
where $T_3$, $T_8$ and $T_{15}$ are the $C$-even non-singlet combinations in the so-called evolution basis, $\Sigma$ the quark singlet and $g$ the gluon distribution, and
$C_i$ is the corresponding Wilson coefficient. Considering a neutron target instead and assuming isospin symmetry, the neutron PDFs are just the same, except for 
$u,\,\bar{u}$ and $d,\,\bar{d}$, which are exchanged. Given the definition of $T_3$, it follows that the neutron structure function $F_2^{n}$ 
can be written with the same Wilson coefficients as in the proton case, and with the only difference given by the sign in front of the $T_3$ contribution:
\begin{align}
    F_2^{n} = C_g \otimes g + C_\Sigma \otimes \Sigma - C_{T_3} \otimes T_3 + C_{T_8} \otimes T_8 + C_{T_{15}} \otimes T_{15} \eqperiod
\end{align}
Considering a deuterium target, which is a bound state of a proton and a neutron, for a generic flavour, the corresponding nuclear 
PDF is the average of a proton and a neutron PDF; hence, the $T_3$ PDF disappears from the expression
\begin{align}
    F_2^{d} = \frac{1}{2} (F_2^p + F_2^n)  = C_g \otimes g + C_\Sigma \otimes \Sigma + C_{T_8} \otimes T_8 + C_{T_{15}} \otimes T_{15} \eqperiod
\end{align}
Thus, one can define the theoretical prediction $G$ for the difference of the proton and deuteron structure functions as
\begin{align}
\label{eq:observable}
    F_2^p - F_2^d = C_{T_3} \otimes T_3 \eqcomma
\end{align}
which is expressed as the integral convolution of the Wilson coefficient $C_{T_3}$ with just one PDF, namely the triplet non-singlet PDF $T_3$.
The determination of $T_3$ using experimental measurements $y$ of $F_2^p - F_2^d$ only involves one flavour and one forward map, making
it an ideal testbed for the method.
This relation is used here in the idealised isoscalar-deuteron approximation, which is appropriate for the reduced closure-test setup. 
Possible deuteron nuclear corrections are not part of the present benchmark but can also be included via an additional term in the covariance matrix~\cite{Ball:2018twp}.
Precomputing the Wilson coefficient and the $T_3$ PDF evolution with the FK tables $W$ discussed in Sec.~\ref{subsec:FKPDF}, and 
identifying $f\equiv xT_3$, since FK tables are conventionally defined to act on the momentum distribution $xf(x)$, we have
\begin{equation}
G_n[f] = \sum_{\alpha = 1}^{N_{\rm grid}}\,
    W_{n,\alpha}\, f(x_\alpha, Q_0^2) \eqcomma
\label{eq:fk-t3}
\end{equation}
where now the sum over flavours has disappeared, as only one single flavour, namely $T_3$, is involved in the forward map for the observable at hand.
The dataset consists of $N_{\rm dat}=248$ measurements of $F_2^p-F_2^d$ in the range $8.75\le Q^2\le 230$~GeV$^2$, 
and the FK table is evaluated on $N_{\rm grid}=50$ points in $x$, spanning $ x\gtrsim 0.07$, corresponding to the lowest $x$ 
measured by the experiments, while the highest value of $x$ is around 0.75. 
This range of $x$ defines the interpolation region used throughout this work, while the region outside it is referred to as the extrapolation region.

The inverse problem is then formulated as the statistical statement that the experimental data
$\mathbf{y} \in \mathbb{R}^{N_{\mathrm{dat}}}$ are related to the {\it unknown} underlying {\it true} triplet PDF
$f^{\ast}$ via
\begin{equation}
  \mathbf{y} = \mathbf{G}[f^{\ast}] + \boldsymbol{\eta} \, \equiv \, \mathbf{y}^{\ast} + \boldsymbol{\eta} \eqcomma
  \qquad \boldsymbol{\eta} \sim \mathcal{N}(\mathbf{0},\, C_{\rm exp}) \eqcomma
\label{eq:inverse-problem}
\end{equation}
with $\boldsymbol{\eta}$ a random vector of experimental noise distributed according to a
Gaussian law with covariance matrix $C_{\rm exp}$. 
The $N_\mathrm{dat} \times N_\mathrm{dat}$ experimental covariance $C_{\rm exp}$ is constructed based on the full experimental covariance matrix 
for the observable $F_2^p - F_2^d$ as
\begin{align}
\label{eq:cmat}
    C_{\rm exp} = \mathrm{Cov}[F_2^p, F_2^p] + \mathrm{Cov}[F_2^d, F_2^d] -2 \mathrm{Cov}[F_2^p, F_2^d] \eqcomma
\end{align} 
thereby retaining all available correlations between proton and deuteron measurements. 
The covariance matrix $C_{\rm exp}$ encodes the statistical and systematic uncertainties of every measurement as well as their correlations
across data points~\cite{Ball:2009qv,Ball:2008by,Ball:2010de}. Moreover, although in the context of a closure test theory uncertainties are not relevant, 
in a global fit to real data it is important to include also the possible sources 
of theory uncertainty, that can be added via a theory covariance matrix $C_{\rm th}$~\cite{NNPDF:2024dpb}. 
The goal of a PDF analysis is to invert Eq.~\eqref{eq:inverse-problem} and to return not a single best-fit PDF, but a probability 
distribution over the space of admissible PDFs that faithfully reflects the uncertainty on the data.
This inversion is an ill-posed problem~\cite{DelDebbio:2021whr,DelDebbio:2026abr}. 
Although the forward map of Eq.~\eqref{eq:fk-t3} is linear, the FK table $W$ acts as a smoothing integral operator: the data constrain $T_3$ only in the range of $x$ covered by the measurements 
and are only weakly sensitive to rapidly varying components of $f$, so that the matrix $W^T C_{\rm exp}^{-1} W$ is ill-conditioned. 
A direct inversion is therefore unstable, and any solution relies on prior assumptions, such as the choice of parametrisation, kernel or regularisation, which in turn shape the 
resulting uncertainty, especially outside the region covered by the data.

\subsection{The closure test approach} 
\label{subsec:closure}
In this work, rather than considering real experimental data $\mathbf{y}$, we work in the framework of a {\it closure test}~\cite{NNPDF:2014otw,DelDebbio:2021whr,NNPDF:2021njg,Harland-Lang:2024kvt}, relying on the NNPDF open-source framework~\cite{zahari_kassabov_2019_2571601,Carrazza:2019mzf,Candido:2022tld,Barontini:2023vmr,Candido:2024rkr}.  
We consider synthetic data generated from a \emph{known} underlying law $f^\ast$, which we set to the central value of the triplet PDF of the NNPDF4.0 global analysis~\cite{NNPDF:2021njg} 
at the parametrisation scale $Q_0=1.65$~GeV. 
The noise-free pseudo-data $\mathbf{y}^\ast=\mathbf{G}[f^\ast]$ are obtained via Eq.~\eqref{eq:fk-t3}, with the $C_{T_3}$ coefficient function computed in perturbative QCD 
at next-to-next-to-leading order (NNLO), while the noise $\boldsymbol{\eta}$ is drawn from the multivariate Gaussian distribution $\mathcal{N}(0,C_{\rm exp})$ centred at zero, 
with $C_{\rm exp}$ defined in Eq.~\eqref{eq:cmat}. 
Hence only the kinematics and the covariance matrix of the BCDMS measurements enter the analysis, while the measured central values do not. 
Using a known underlying law allows us to assess how well each methodology reconstructs the \emph{truth}, and whether its uncertainty is compatible with the actual deviation from it.

In the context of a closure test, the $\mathbf{y}^\ast$ vector of Eq.~\eqref{eq:inverse-problem}, namely the vector of 
synthetic data generated out of the {\it true} underlying distribution, is often referred to as Level-0 ($L_0$) data. The 
$\mathbf{y}$ vector of Eq.~\eqref{eq:inverse-problem}, namely the {\it true} data plus some randomly drawn noise $\boldsymbol{\eta}$, is 
referred to as Level-1 ($L_1$) data.
In a Monte Carlo replica sampling of the probability density~\cite{Costantini:2024wby}, one also introduces a Level-2 ($L_2$) data level, which is 
generated by augmenting $\mathbf{y}$ with a replica-dependent noise $\boldsymbol{\epsilon}^{(k)}$
\begin{align}
   \mathbf{y}^{(k)} &= \mathbf{y} + \boldsymbol{\epsilon}^{(k)} = \mathbf{y}^* + \boldsymbol{\eta} +  \boldsymbol{\epsilon}^{(k)}, \quad k=1,\ldots,N_{\rm rep} \eqcomma
    \label{eq:l2_pseudodata}
\end{align}
where $k$ is the replica index and $N_{\rm rep}$ is the number of synthetic data replicas that are generated. 
Each instance of the noise $\boldsymbol{\epsilon}^{(k)}$ is drawn independently from the same
multi-Gaussian distribution $\mathcal{N}(0, C_{\rm exp})$, which the
observational noise is drawn from. This implies that 
the covariance of any two data points over the replica sample in the limit of large $N_{\rm rep}$ tends to the corresponding $C_{\rm exp}$ matrix element.
The interpretation of the spread of an ensemble of fits at each level is discussed in Sec.~\ref{sec:unc_quant}.

\section{Architectures and Uncertainties}
\label{sec:setup}
In this section, we describe the regression strategies used to reconstruct the triplet PDF function $T_3(x,Q_0^2)$ from data-space constraints.
We compare GPs, NNs and NTK approximations.
All methods are applied to the same inverse problem, as discussed in Sec.~\ref{sec:t3model}.
The purpose of this section is twofold.
First, we define the architectures, loss functions and preprocessing choices used in the numerical comparison.
Second, we specify how uncertainties are estimated for each method, in particular the distinction between the experimental covariance $C_{\rm exp}$ and the ensemble or PDF covariance $C_{\rm ens}$.

Throughout this section, we denote by $f_\theta(x)\equiv xT_3(x,Q_0^2)$ the triplet PDF, multiplied by $x$, that is inferred from the data, where $\theta$ are the parameters 
characterising it (whether parametrised via NNs or GPs), 
and their distribution determines the probability density of $f$ inferred from the data. 
The observable-space prediction obtained from $f_\theta$ is
\begin{align}
    \hat G_\theta = W f_{\theta} \eqcomma
    \label{eq:discrete-forward-map}
\end{align}
where $W$ is the forward map introduced in Sec.~\ref{sec:t3model}, and $\hat G_\theta$ is the model prediction for the observables $\mathbf{y}$ of Eq.~\eqref{eq:inverse-problem}. 
While all likelihoods and loss functions are evaluated in observable space, we present our final results in both PDF and, where relevant, observable spaces.

Throughout, $x$ denotes the momentum fraction and $u=\log x$ (with $u'=\log x'$) the logarithmic input variable used by all architectures. 
Functions and kernels are labelled by $x$, e.g.\ $f(x)$ and $k(x,x')$, with the logarithmic transformation of the inputs understood, 
while the endpoint prefactors of Eq.~\eqref{eq:scaling} are always evaluated at $x$.

For each architecture, since we have theoretical input about the power behaviour of the triplet PDF at small- and large $x$, it is convenient to
encode such input in the prior. We do so by including endpoint preprocessing, which introduces controlled small-$x$ and large-$x$ behaviour 
through a multiplicative prefactor, 
\begin{align}
    x^\alpha(1-x)^\beta \eqcomma
    \label{eq:scaling}
\end{align}
with $\alpha$ and $\beta$ treated as free parameters, trained together with the network weights for the NN and sampled with the kernel hyperparameters for the GP (Sec.~\ref{sec:GP}). 
In the trainable-preprocessing setups, the exponents are constrained to the ranges of $\alpha\in[0.1,1.1]$ and, respectively, $\beta\in[1.7,3.5]$.
These bounds allow moderate flexibility in the small-$x$ behaviour while ensuring that $T_3\propto x^{\alpha-1}$ is integrable at $x=0$, 
with the expected Regge behaviour $T_3\sim x^{-1/2}$ within the allowed range, and that $T_3$ vanishes at $x=1$, while avoiding extreme endpoint shapes. 
This range of $\alpha$ is essentially equivalent to the one adopted in Ref.~\cite{Candido:2024hjt}, where the prefactor $x^{\alpha'}$ multiplies $T_3$ with $\alpha'\in(-1,0)$, 
corresponding to $\alpha=\alpha'+1\in(0,1)$ in our convention. 
In practice, the trained values remain close to their initial values, so the preprocessing acts mainly as a controlled functional bias rather 
than as an additional highly flexible parametrisation.
In what follows we use the same initial values, $\alpha=1.0$ and $\beta=3.0$, for the NN and NTK setups, while for the GP they are sampled from uniform priors over the same ranges. 
The choice $\alpha=1.0$ corresponds to a flat small-$x$ behaviour of $T_3$, rather than the Regge-like behaviour $T_3\sim x^{-1/2}$ ($\alpha\simeq0.5$). 
It is motivated by the GP posteriors of App.~\ref{app:scan} and by previous work~\cite{Candido:2024hjt}, but should be regarded as a modelling assumption. 
Since the trained exponents remain close to their initial values, the small-$x$ extrapolation and its uncertainty are conditional on this choice: 
as shown in Fig.~\ref{fig:alpha_scan_fixed_beta}, fits with $\alpha=0.5$ and $\alpha=1.0$ agree in the data region but differ substantially below $x\simeq0.07$. 
Additional scans over fixed and trainable values of $\alpha$ and $\beta$, including fixed values outside the default range, are discussed in App.~\ref{app:scan}. 

\subsection{Gaussian process}
\label{sec:GP}
A GP is a non-parametric supervised learning method for regression and probabilistic classification tasks~\cite{rasmussen2006gaussian}.
For regression, it defines a probability distribution over functions, thereby providing not only a central prediction but also an associated uncertainty estimate. 
To account for different behaviours in the small-$x$ and large-$x$ regions, the distances between two points entering the kernels below are measured in $u$.

The GP combines a prior distribution over functions with a likelihood based on the training data. 
We can write the prior as
\begin{align}
    f_\theta(x) \sim \mathcal{GP}\bigl(m(x),k(x,x')\bigr) \eqcomma
\end{align}
where $m(x)$ denotes the prior mean and $k(x,x')$ the covariance kernel, both of which depend on a set of hyperparameters $\theta$. 
In this work, we use a zero prior mean, $m(x)=0$.  
The covariance kernel determines the correlation between function values at different input points and therefore controls the smoothness of the functions drawn from the prior.
\newline
For a finite set of grid points
\begin{align}
    X = \{x_\alpha\}_{\alpha=1}^{N_{\rm grid}}\; ,
\end{align}
the GP prior becomes a multivariate Gaussian,
\begin{align}
    f_X \equiv \bigl(f(x_1),\ldots,f(x_{N_{\rm grid}})\bigr)^T \sim \mathcal{N}\bigl(m_X,K_{XX}\bigr) \eqcomma
\end{align}
where the covariance matrix is constructed by evaluating the kernel on all pairs of grid points,
\begin{align}
    (K_{XX})_{ij}=k(x_i,x_j) \eqperiod
\end{align}
The key idea is that if two points $x_i$ and $x_j$ are considered similar by the kernel, then the corresponding function values $f(x_i)$ and $f(x_j)$ are expected to be similar as well.

In the present inverse problem, the likelihood is defined for the data vector $y$, not directly for the PDF values $f_X$.
Using the forward map of Eq.~\eqref{eq:discrete-forward-map}, we can write the likelihood as
\begin{align}
    p(y|f_X)=\mathcal{N}\bigl(y\,\big|\,W f_X,C_{\rm exp}\bigr) \eqperiod
\end{align}
Combining this likelihood with the GP prior yields the posterior distribution over PDFs,
\begin{align}
    p(f_X|y)\propto p(y|f_X)\,p(f_X) \eqperiod
\end{align}
When endpoint preprocessing is enabled for the GP, the prior is placed on an auxiliary latent function $h(x)$, and the PDF is defined as
\begin{align}
    f(x)=x^\alpha(1-x)^\beta h(x) \eqperiod
\end{align}
Since the latent function $h(x)$ is Gaussian and sign-unconstrained, the resulting $T_3$ prior does not impose pointwise positivity.
Thus, based on this relation, we can induce a preprocessed covariance for $f$,
\begin{align}
    k_f(x,x')=x^\alpha(1-x)^\beta\, k_h(x,x')\, x'^\alpha(1-x')^\beta \eqperiod
\end{align}
The kernels of Eqs.~\eqref{eq:rbf_kernel}--\eqref{eq:matern_kernel} below specify the latent covariance $k_h$, while $k$ in the GP prior above denotes the full covariance $k_f$, 
which coincides with $k_h$ when no preprocessing is applied.
This implements the same endpoint structure used in the NN setup while keeping the likelihood in observable space.
For GP fits with endpoint preprocessing, $\alpha$ and $\beta$ are treated as kernel hyperparameters and sampled together with the remaining GP hyperparameters during marginal-likelihood inference.

We test three different kernels in the GP implementation: a radial basis function (RBF) kernel, a Gibbs kernel and a Mat\'ern kernel. 
The RBF kernel, also known as the squared-exponential or Gaussian kernel, is written as~\cite{rasmussen2006gaussian}
\begin{align}
    k_{\rm RBF}(x,x')=\,\sigma^2
    \exp\left[-\frac{(u-u')^2}{2\ell^2}\right] \eqcomma
    \label{eq:rbf_kernel}
\end{align}
where $\ell$ denotes the kernel length scale and $\sigma$ is a constant prefactor. 
\newline
Based on the RBF kernel, the Gibbs kernel introduces an $x$-dependent length
scale~\cite{Gibbs:1997GP,Paciorek:2004NonstationaryGP},
\begin{align}
    k_{\rm Gibbs}(x,x')=\sigma^2\sqrt{\frac{2\ell(x)\ell(x')}{\ell^2(x)+\ell^2(x')}}\exp\left[-\frac{(u-u')^2}{\ell^2(x)+\ell^2(x')}\right] \eqcomma
    \label{eq:gibbs_kernel}
\end{align}
where the length scale $\ell(x) = \ell_0(x+\delta)$ is a function of $x$, while the distance it controls is measured in $u$, with $\ell_0$ a tunable hyperparameter and $\delta>0$ ensuring a positive length scale.
The Gibbs kernel is non-stationary and therefore allows the smoothness of the GP prior to vary across the $x$-range.

Lastly, we consider the Mat\'ern kernel, a generalisation of the RBF kernel which contains an additional parameter $\nu$, controlling the smoothness of the
sampled functions~\cite{rasmussen2006gaussian}. 
It is defined as
\begin{align}
    k_{\rm Mat\acute ern}(x,x')=\sigma^2\frac{1}{\Gamma(\nu)2^{\nu-1}}\left(\frac{\sqrt{2\nu}}{\ell} |u-u'|\right)^\nu K_\nu\left(\frac{\sqrt{2\nu}}{\ell} |u-u'|\right) \eqcomma
    \label{eq:matern_kernel}
\end{align}
where $|u-u'|$ is the distance between two points in $u=\log x$, $K_\nu$ the modified Bessel function, and $\Gamma$ the Gamma function. 
In the limit $\nu \to \infty$, the Mat\'ern kernel approaches the RBF kernel, while smaller values of $\nu$ allow for less smooth functions. In this study, we use $\nu=\{0.5,1.5,2.5\}$ to test different smoothness assumptions, ranging from the exponential kernel at $\nu=0.5$ to smoother Mat\'ern priors at larger $\nu$.

For a fixed kernel family, we infer the GP hyperparameters $\theta$ by sampling their posterior distribution, with prior ranges listed in App.~\ref{app:hyper} 
and likelihood given by the GP marginal likelihood, which for the linear forward map of Eq.~\eqref{eq:discrete-forward-map} reads
\begin{align}
    p(\mathbf{y}|\theta)=\mathcal{N}\bigl(\mathbf{y}\,\big|\,W m_X,\;W K_{XX} W^T + C_{\rm exp}\bigr) \eqcomma
    \label{eq:gp-marginal}
\end{align}
where $K_{XX}$ (and, in general, $m_X$) depend on $\theta$. 
The full experimental covariance matrix therefore enters the GP inference exactly.
The posterior predictive replicas are then generated by looping over the hyperparameter samples $\theta^{(s)}$, constructing the corresponding conditional GP posterior
\begin{align}
    p(f_{\rm pred}|\mathbf{y},\theta^{(s)}) = \mathcal{N}\!\left(m_\star(\theta^{(s)}),K_\star(\theta^{(s)})\right) \eqcomma
\end{align}
and drawing one function replica from each conditional posterior. 
The empirical covariance of these replicas is used as the GP contribution to $C_{\rm ens}$. 
We treat different kernel families as different model specifications rather than combining them into a single Bayesian model average.

\subsection{Neural Network}
\label{sec:NN}
As the underlying architecture for the NN, we always use a multilayer perceptron (MLP), which has been widely used in PDF determinations for its flexibility as a universal function approximator~\cite{DelDebbio:2007ee,Ball:2008by,Ball:2017nwa}.
As for the GP, the input variable is $u=\log x$, linearly rescaled to the training interval to improve the resolution of the small-$x$ region, and the network output is the PDF-space prediction $xT_{3,\theta}(x,Q_0^2) \equiv f_\theta(x)$, where $\theta$ denotes the trainable parameters.
Its data-space prediction is the architecture-specific instance of Eq.~\eqref{eq:discrete-forward-map}. 
The loss is therefore evaluated in data space by comparing $\hat G_\theta$ to the training data $y$, rather than comparing $f_\theta$ directly to the underlying closure-test truth.

The NNs consist of an input layer followed by $n$ hidden blocks and an output layer.
Each hidden block contains a fully connected layer, a non-linear activation function and a dropout layer~\cite{Srivastava:2014Dropout}.
In the numerical implementation, we use the GELU activation function~\cite{Hendrycks:2016GELU}.
As discussed at the beginning of this section, we also apply the endpoint preprocessing $x^\alpha(1-x)^\beta$, with $\alpha$ and $\beta$ treated as trainable hyperparameters.
Writing $N_\theta(u)$ for the raw network output, the final prediction is
\begin{align}
    f_\theta(x)=x^\alpha(1-x)^\beta\,N_\theta(u) \eqperiod
    \label{eq:nn-output}
\end{align}
No positivity constraint is applied to the final $T_3$ output, as this particular PDF combination does not have to be positive~\cite{Candido:2021zcp}. 
All details about the NN's architecture and training hyperparameters are given in Appendix~\ref{app:hyper}.

We consider three different loss functions. 
The selection of a loss function is inherently dictated by the nature of the data and the availability of uncertainty information. 
In traditional PDF fitting, the standard approach is to optimise a log-likelihood loss that explicitly incorporates the experimental 
covariance matrix $C_{\rm exp}$. 

This formulation is essential because it accounts for the full estimate of systematic uncertainties and point-to-point correlations present 
in experimental data. However, as ML models are increasingly deployed for tasks beyond standard fitting, such as constructing surrogate models for complex scattering amplitudes~\cite{Bahl:2024gyt, Bahl:2025xvx}, an experimental covariance matrix is often unavailable. In these simulation-driven regimes, a 
standard mean-squared error (MSE) loss is typically used to optimise the central prediction:
\begin{align}
\mathcal{L}_\mathrm{MSE}= \frac{1}{N_{\rm dat}}\sum_{n=1}^{N_{\rm dat}}\left(\hat G_{\theta,n}-y_n\right)^2 \eqcomma
\label{eq:MSE_loss}
\end{align}
where $N_{\rm dat}$ represents the number of training data points. Mirroring the strategy used in that context to quantify uncertainties 
without an experimental error budget, we also consider a heteroscedastic loss. This allows the model to dynamically learn 
and attribute data-dependent variance to the outputs by predicting both a central value and a data-point-dependent variance, denoted by $\sigma_{\theta,n}^2$. 
The corresponding heteroscedastic loss is defined as:
\begin{align}
\mathcal{L}_\mathrm{het}=\frac{1}{N_{\rm dat}}\sum_{n=1}^{N_{\rm dat}}\left[\frac{\left(\hat G_{\theta,n}-y_n\right)^2}{2\sigma_{\theta,n}^2}
+\frac{1}{2}\log \sigma_{\theta,n}^2\right] \eqcomma
\label{eq:het_loss}
\end{align}
where $\sigma_{\theta,n}^2$ is the learned diagonal variance in data space. This formulation corresponds to a Gaussian negative log-likelihood 
assuming a diagonal, input-dependent covariance structure. Consequently, while it can capture pointwise variances, it cannot encode correlated 
experimental uncertainties. Our goal is to assess whether this heteroscedastic loss can accurately infer the underlying uncertainty of the training data. 
We benchmark its performance against a standard log-likelihood optimisation (hereafter termed the $\chi^2$ loss) evaluated on the $L_0$, $L_1$, or $L_2$ data vectors. 
The $\chi^2$ loss, which fully incorporates the experimental covariance matrix, reads:
\begin{align}
\mathcal{L}_{\chi^2}=\left(\hat G_\theta-y\right)^\top C_{\rm exp}^{-1}\left(\hat G_\theta-y\right) \eqcomma
\label{eq:chi2_loss}
\end{align}
where $C_\mathrm{exp}$ is the covariance matrix of the data entering the fit, as introduced in Sec.~\ref{sec:t3model}. 
In NNPDF fits, the $t_0$ prescription is used to avoid the bias associated with multiplicative normalisation uncertainties~\cite{Ball:2009qv}; 
here we use $C_{\rm exp}$ throughout, and show in App.~\ref{app:L2} that adopting $C_{t_0}$ does not alter our conclusions. 
In summary, the MSE loss given in Eq.~\eqref{eq:MSE_loss} weights all data points equally; the heteroscedastic loss in 
Eq.~\eqref{eq:het_loss} adapts these weights in an uncorrelated, pointwise manner; and the $\chi^2$ loss in Eq.~\eqref{eq:chi2_loss} 
fully accounts for correlated experimental uncertainties. The $\chi^2$ approach closely mirrors standard PDF fitting methodologies, 
where the agreement between theory predictions and data is rigorously measured via the experimental covariance matrix~\cite{Ball:2008by,Ball:2017nwa}. 

Finally, because a single deterministic NN provides only a point estimate, we determine the model spread by training an ensemble of 
independent networks. Depending on the configuration, the ensemble variation originates from random initialisations, stochastic optimisation, 
and—for $L_2$-based fits—distinct data replicas. The empirical covariance of these ensemble predictions defines the NN contribution 
to $C_{\rm ens}$, as detailed in Sec.~\ref{sec:unc_quant}.
Overall, the ensemble of NNs trained on the $L_2$ data should be understood as a simplified, NNPDF-inspired construction 
rather than a full implementation of the NNPDF fitting methodology.
It reproduces the basic idea of fitting independently fluctuated data replicas, but does not include all ingredients of a full NNPDF global analysis, 
such as multi-flavour refitting, dataset-by-dataset treatment, cross-validation prescriptions, positivity constraints, or a full methodological optimisation.

\subsection{Neural Tangent Kernel}
\label{sec:NTK}
Next, we introduce the Neural Tangent Kernel (NTK), which provides a kernel-based description of the training dynamics of a NN. 
Originally derived to describe the evolution of NNs under gradient descent in the infinite-width limit~\cite{Jacot:2018NTK}, the NTK should 
be distinguished from the neural-network Gaussian process (NNGP) kernel. While the NNGP kernel characterises the prior distribution 
over functions at initialisation~\cite{Neal:1996nn, Lee:2018DeepGP}, the NTK governs the linearised training dynamics during optimisation. 

For a network output $f_\theta(x)$ parametrised by trainable parameters $\theta$, including weights, biases and the endpoint exponents, the empirical NTK is defined as
\begin{align}
k_{\rm NTK}(x,x';\theta)=\nabla_\theta f_\theta(x)\cdot\nabla_\theta f_\theta(x') \eqperiod
\label{eq:ntk}
\end{align}
Here, $f_\theta(x)$ denotes the final, sign-unconstrained PDF-space network output, incorporating the endpoint prefactor evaluated at $x$ and the logarithmic input $u=\log x$.
The kernel thus measures the similarity between two inputs $x$ and $x'$ via the inner product of the network's parameter gradients. 
Crucially, unlike the analytical kernels chosen externally for standard GPs, the NTK is intrinsically induced by 
the architecture, activation functions, and initialisation scheme of the underlying NN. 

In this work, the reference network is the NN described in Sec.~\ref{sec:NN} (with hyperparameters detailed in 
App.~\ref{app:hyper}), including identical endpoint preprocessing where enabled. During training, the network parameters 
evolve, causing the empirical NTK to change over time. However, in the infinite-width limit under appropriate 
parametrisation, the NTK becomes a deterministic, static object throughout optimisation~\cite{Jacot:2018NTK, Lee:2019WideNN, Chiefa:2025cap}. 
This corresponds to the so-called ``lazy-training'' regime, where individual parameters move only infinitesimally from their initial values 
$\theta_0$ while their collective updates yield a finite change in the network output. 
In this regime, the network can be linearised via a first-order Taylor expansion around initialisation:
\begin{align}
f(x;\theta_0+\Delta\theta)\simeq f(x;\theta_0)+\Delta\theta\cdot\nabla_\theta f(x;\theta_0) \eqcomma
\label{eq:ntk-linearisation}
\end{align}
where we write $f(x;\theta) \equiv f_\theta(x)$ to make the parameter dependence explicit. Under this 
linearisation, training the network with gradient descent is exactly equivalent to kernel gradient descent 
using $k_{\rm NTK}$. 

In our numerical comparisons, we do not assume the strict infinite-width limit. Instead, we use the empirical NTK 
as a local finite-width diagnostic to evaluate the local tangent model at selected stages of the non-linear training trajectory, 
using the empirical NTK introduced in Ref.~\cite{Chiefa:2025cap}. 
At a given training epoch $s$ with parameters $\theta_s$, the linearised model is expressed as
\begin{align}
f_{\rm lin}^{(s)}(x) = f_{\theta_s}(x) + \nabla_\theta f_{\theta_s}(x)\cdot \delta\theta_s \eqcomma
\label{eq:ntk-stage-linearisation}
\end{align}
where $f_{\theta_s}(x)$ is the frozen non-linear network prediction at stage $s$, and $\delta\theta_s$ is a displacement in parameter space. 
Assigning a Gaussian prior $\delta\theta_s\sim\mathcal{N}(0,\mathbb{I})$ to the displacement induces a GP prior on the linearised model,
\begin{align}
f_{\rm lin}^{(s)}(x)\sim\mathcal{GP}\big(f_{\theta_s}(x),\,k_{\rm NTK}(x,x';\theta_s)\big) \eqcomma
\label{eq:ntk-gp}
\end{align}
whose mean is the network output and whose covariance is the empirical NTK of Eq.~\eqref{eq:ntk} evaluated at $\theta_s$. 
We condition this GP on the data with the same Gaussian likelihood used for the GPs in Sec.~\ref{sec:GP}, which includes the full experimental covariance $C_{\rm exp}$. 
The posterior mean provides the NTK central prediction and is equivalent to minimising the $\chi^2$ loss of Eq.~\eqref{eq:chi2_loss} for the linearised model 
of Eq.~\eqref{eq:ntk-stage-linearisation} with a quadratic penalty on $\delta\theta_s$, while the posterior covariance provides the NTK uncertainty estimate. 
This construction provides a natural bridge between the implicit features learned by a NN and the explicit functional prior of a GP. 

Our construction differs from the gradient-flow analysis of Ref.~\cite{Chiefa:2025cap}. There, the empirical NTK is approximated as constant after an initial training transient, allowing the subsequent dynamics to be solved analytically. In the eigenbasis of the resulting training operator, components approach their stationary values with factors $e^{-h_i t}$, where the rates $h_i$ depend jointly on the NTK, the forward map and the experimental covariance. Finite training time therefore suppresses slowly relaxing components and acts as a regularisation parameter. Here, at each selected epoch $s$, we use the network output and empirical NTK to define a GP prior and obtain predictions and uncertainties by conditioning on the data. The posterior depends on $s$ through its prior mean and kernel, without introducing an additional gradient-flow evolution.

We evaluate this linear NTK approximation at three distinct training stages: $s=0$ (random initialisation), $s=100$, and $s=5000$ 
(the maximum optimisation epoch). While at the initialisation stage ($s=0$) the kernel and prior mean are independent of the data, the latter two stages utilise parameter 
configurations reached via the non-linear $\chi^2$-trained NN, thereby probing the localised geometry of the trained network along its optimisation trajectory. 
The stage $s=100$ is chosen to probe the early phase of training: as shown in Sec.~\ref{sec:results}, the NTK evaluated after 100 epochs already captures 
most of the improvement in the central prediction, while further training mainly affects the uncertainty in the extrapolation region. 
For the NNPDF architecture, Ref.~\cite{Chiefa:2025cap} showed that the empirical NTK of the $T_3$ fit becomes stationary after an initial transient. 
We do not repeat this analysis for the architecture and training setup used here, and therefore do not assume that the kernel at $s=100$ or $s=5000$ is stationary.
In all three cases, the GP is conditioned on the same data with the same covariance $C_{\rm exp}$; the stages differ only in the network parameters at which the kernel and the prior mean are evaluated. 
At finite width, the empirical NTK also depends on the random initialisation and on the optimisation path. 
Since we use a single network, this variability is not included in the NTK uncertainty estimate.


\subsection{Uncertainty decomposition and diagnostics}
\label{sec:unc_quant}
In physics analyses, uncertainties are commonly separated into statistical and systematic components~\cite{Cowan:1998StatisticalDataAnalysis,Haussmann:2026gbi}, the former arising from finite data samples and stochastic fluctuations and therefore decrease as more data become available, the latter arising from imperfect knowledge of the experimental conditions, detector effects, or theoretical modelling assumptions.
In the machine-learning and statistics literature, uncertainties are often classified as aleatoric and epistemic~\cite{Kendall:2017Uncertainties,Gal:2016Uncertainty}.
Aleatoric uncertainties are data uncertainties associated with the intrinsic randomness of the data-generating process. 
Epistemic uncertainties are model uncertainties associated with limited knowledge of the model, its parameters, or its functional form. 
These two classifications are complementary rather than identical, where aleatoric and epistemic uncertainties refer to the origin of the uncertainty, while statistical and systematic uncertainties refer to their behaviour under repeated measurements or improved modelling. 
In the numerical results of this work, we use a simpler operational separation into two contributions:
\begin{itemize}
    \item \textbf{Data uncertainty.}
    The first contribution is the experimental covariance matrix, denoted by $C_{\rm exp}$, which encodes the uncertainty of the data entering the fit.
    This contribution is associated with the aleatoric, data-driven part of the uncertainty quantification.
    \item \textbf{Model uncertainty.}
    The second contribution is referred to as PDF uncertainty, and is given by the ensemble covariance matrix, denoted by $C_{\rm ens}$, which is 
    obtained from the spread of independent model predictions. 
    This contribution captures model dependence induced by finite and noisy training data, random initialisations, stochastic optimisation, 
    posterior sampling, or, in the case of NTKs, the GP posterior induced by the empirical kernel. 
    We therefore use $C_{\rm ens}$ as a practical estimator of the epistemic component, 
    with the caveat that, for $L_2$ ensembles, $C_{\rm ens}$ also contains the data uncertainty propagated to PDF space.
\end{itemize}
For an ensemble of $N_{\rm rep}$ predictions $f_k(x)$, with $k=1,\ldots,N_{\rm rep}$, the ensemble mean is
\begin{align}
    \bar f(x)=\frac{1}{N_{\rm rep}}\sum_{k=1}^{N_{\rm rep}} f^{(k)}(x) \eqcomma
\end{align}
and the corresponding ensemble covariance is defined as
\begin{align}
    C_{\rm ens}^{(f)}(x_i,x_j)=\frac{1}{N_{\rm rep}-1}\sum_{k=1}^{N_{\rm rep}}\left[f^{(k)}(x_i)-\bar f(x_i)\right]\left[f^{(k)}(x_j)-\bar f(x_j)\right] \eqperiod
    \label{eq:cens}
\end{align}
The same definition can be applied in observable space after applying the forward map.

\begin{figure}[b!]
    \centering
    \includegraphics[width=0.49\textwidth, page=1]{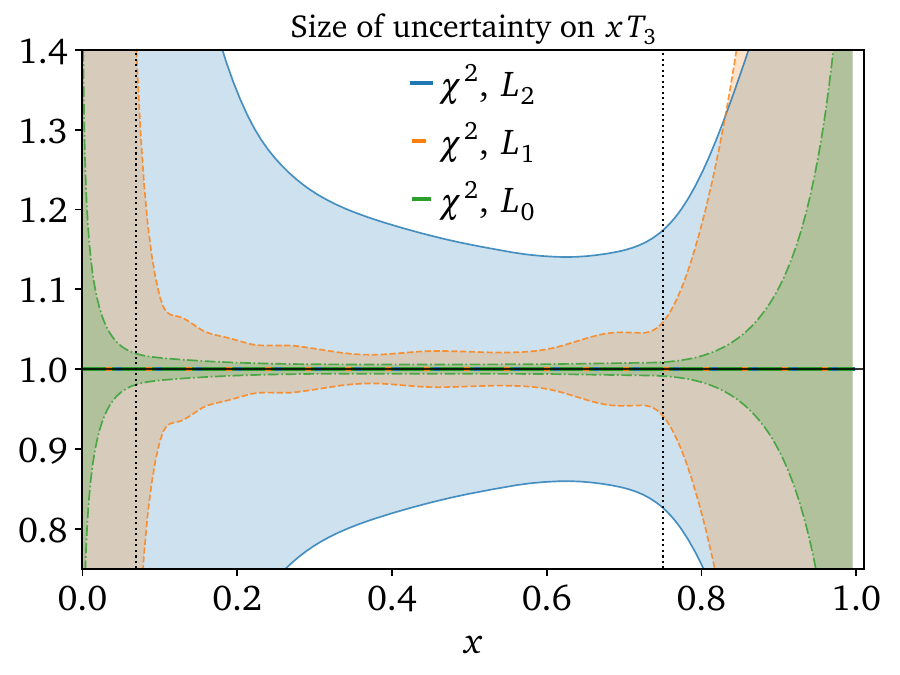}
    \includegraphics[width=0.49\textwidth, page=2]{figs/xt3_truth_ratio.pdf}
    \caption{Comparison of the normalised uncertainties $\sigma_{\rm norm}$ and deviations from the truth for a full NN setup with a $\chi^2$ loss using either $L_0$, $L_1$ or $L_2$ data. The vertical lines indicate the distinction between interpolation and extrapolation regions.}
    \label{fig:unc_bands}
\end{figure}

For NNs, $C_{\rm ens}$ is determined from an ensemble of independently trained networks. 
Following Refs.~\cite{NNPDF:2014otw,DelDebbio:2021whr,NNPDF:2021njg}, the spreads of ensembles trained at the three levels of closure-test data 
defined in Sec.~\ref{subsec:closure} are cumulative and can be interpreted as follows:
\begin{itemize}
    \item \textbf{Level-0 ($L_0$):} with noise-free data, the ensemble spread measures the interpolation and extrapolation uncertainty, i.e.\ the freedom left by a finite set of data points, together with residual training variations.
    \item \textbf{Level-1 ($L_1$):} fitting a single noisy data realisation adds a functional uncertainty, reflecting the existence of different functions that describe the same noisy data equally well.
    \item \textbf{Level-2 ($L_2$):} fitting independent data replicas, Eq.~\eqref{eq:l2_pseudodata}, further adds the propagated data uncertainty, so that the $L_2$ spread provides the total PDF uncertainty.
\end{itemize}
In Fig.~\ref{fig:unc_bands} we show a comparison of the normalised uncertainties
\begin{align}
    \sigma_{\rm norm}(x) = \frac{\sqrt{C_{\rm ens}^{(f)}(x,x)}}{|\bar f(x)|} \eqcomma
\end{align}
for an ensemble of NNs with a $\chi^2$ loss trained on $L_0$, $L_1$ and $L_2$ data.
Here, $\bar f(x)$ is the ensemble mean of $f_\theta(x)\equiv xT_3(x,Q_0^2)$, and $C_{\rm ens}^{(f)}$ is the corresponding ensemble covariance.
The shaded bands show $1 \pm \sigma_{\rm norm}(x)$.
We observe that in the data region the $L_0$ component of the PDF uncertainty is the smallest, while it becomes visible and significant in the extrapolation region. 
The $L_2$ uncertainty component dominates the data region and the $L_1$ is highly enhanced in the extrapolation region. 

For GPs, the uncertainty entering $C_{\rm ens}$ is obtained from the empirical spread of posterior predictive samples for a fixed GP setup. 
For a given kernel family, this spread includes the conditional GP posterior uncertainty and the dependence on sampled kernel hyperparameters. 
Different kernel families are compared as separate model choices rather than combined into a single covariance.

For the NTK, $C_{\rm ens}$ is the posterior covariance of the GP defined by the empirical NTK of a single network, Eq.~\eqref{eq:ntk-gp}, 
evaluated at the three training stages discussed in Sec.~\ref{sec:NTK}. 
Since the GP posterior already provides a distribution over functions, no additional ensemble of networks is used, 
and $C_{\rm ens}$ does not include the variability of the empirical NTK across random initialisations.

To test whether the ensemble uncertainty estimate is statistically compatible with the observed residuals, we use a normalised-bias diagnostic.
We consider the different ensemble members $k$ of the chosen architecture, either the NN, NTK or GP, by averaging the individual 
predictions $\hat G^{(k)}$ towards a mean prediction 
\begin{align}
    \hat{\bar G} = \frac{1}{N_{\rm rep}}\sum_{k=1}^{N_{\rm rep}} \hat G^{(k)} \eqperiod
\end{align}
Since $C_{\rm ens}$ in data space is obtained from predictions $\hat G^{(k)}=Wf^{(k)}$ defined on $N_{\rm grid}=50$ points, its rank is at most $N_{\rm grid}<N_{\rm dat}$, 
and it cannot be inverted directly. 
We therefore decompose it into eigenvectors, $C_{\rm ens}=\sum_{i}\lambda_i\,\mathbf{v}_i\mathbf{v}_i^\top$ with $\lambda_1\ge\lambda_2\ge\ldots$, and, following the procedure of Ref.~\cite{Barontini:2025lnl} with a stricter criterion, 
retain only the $n_{\rm eig}$ eigenvectors with the largest eigenvalues, such that the trace of the regularised matrix differs by at most 0.1\% from that of the original one. 
The inverse of $C_{\rm ens}$ is then replaced by its pseudo-inverse on the retained subspace,
\begin{align}
    C^{+}_{\rm ens}=\sum_{i=1}^{n_{\rm eig}}\frac{\mathbf{v}_i\mathbf{v}_i^\top}{\lambda_i} \eqcomma
\end{align}
and we define the normalised bias $R_b$ with respect to the underlying theory $\mathbf{y}^\ast$ as
\begin{align}
    R_b=\sqrt{\frac{1}{n_{\rm eig}}\left(\hat{\bar G}-\mathbf{y}^\ast\right)^\top C^{+}_{\rm ens} \left(\hat{\bar G}-\mathbf{y}^\ast\right)} \eqperiod
    \label{eq:normalised-bias}
\end{align}
The normalisation by $n_{\rm eig}$, the number of retained eigenvalues, ensures that $R_b^2$ has unit expectation value if the deviation of the ensemble mean from the truth 
is distributed according to $C_{\rm ens}$. 
The same regularisation and normalisation are applied when $C_{\rm ens}$ is replaced by its diagonal, $\mathrm{diag}(C_{\rm ens})$, 
in which case the retained eigenvectors correspond to the data points with the largest variances.
For a faithful uncertainty, $n_{\rm eig}R_b^2$ follows a $\chi^2$ distribution with $n_{\rm eig}$ degrees of freedom, 
so that on a single realisation $R_b$ fluctuates around one with a standard deviation of approximately $1/\sqrt{2n_{\rm eig}}$. 
For example, for $n_{\rm eig}=7$ values of $R_b$ between $0.49$ and $1.51$ are compatible with a faithful uncertainty at the 95\% confidence level. 
When the diagonal approximation is used, $n_{\rm eig}$ is much larger, but since the residuals are strongly correlated, $n_{\rm eig}R_b^2$ no longer follows a $\chi^2$ distribution with $n_{\rm eig}$ degrees of freedom, 
and the statistical precision is not improved correspondingly.
The normalised bias $R_b$ is a diagnostic measure of whether the deviation of the ensemble mean prediction from the underlying truth $\mathbf{y}^\ast$ 
is compatible with the ensemble covariance $C_{\rm ens}$.
It is not intended as a test of whether the generated $L_1$ pseudo-data themselves follow $C_{\rm exp}$, which is fixed by construction.
The diagnostic is evaluated only in the interpolation region, where the data-space covariance, the FK map, and the reference data are defined.
If this deviation is statistically compatible with $C_{\rm ens}$, one expects $R_b\simeq 1$. 
$R_b$ is the single-realisation analogue of the bias--variance ratio of Refs.~\cite{DelDebbio:2021whr,NNPDF:2021njg}: 
since it is evaluated on a single $L_1$ realisation, values of order one, rather than exactly one, are expected for a faithful uncertainty.
Values $R_b>1$ indicate that the covariance underestimates the observed deviation, while values $R_b<1$ indicate that the 
chosen covariance is too large for the observed deviation, or that it contains directions with excessive variance.
\section{Results}
\label{sec:results}
In this section, we compare the performance of different regression strategies introduced in Sec.~\ref{sec:setup}.
In Sec.~\ref{sec:arc_res} we study their ability to reconstruct the underlying $T_3$ distribution in a closure-test 
setting, both inside the data-supported interpolation region and in the extrapolation region.
We then test the faithfulness of the corresponding uncertainty estimates in two complementary ways.
In Sec.~\ref{sec:bias_res} the normalised bias of Eq.~\eqref{eq:normalised-bias} tests the uncertainties in the data 
(interpolation) region, $x\gtrsim0.07$, in which the models are trained.
In Sec.~\ref{sec:hera_res} we test them in the extrapolation region, by propagating the reconstructed $T_3$ to HERA 
reduced cross sections, which partly lie at values of $x$ below those covered by the BCDMS data and therefore probe 
a region that was not used to train the models.

\subsection{Central reconstruction and extrapolation behaviour}
\label{sec:arc_res}
\begin{figure}[t!]
    \centering
   \includegraphics[width=0.6\textwidth, page=1]{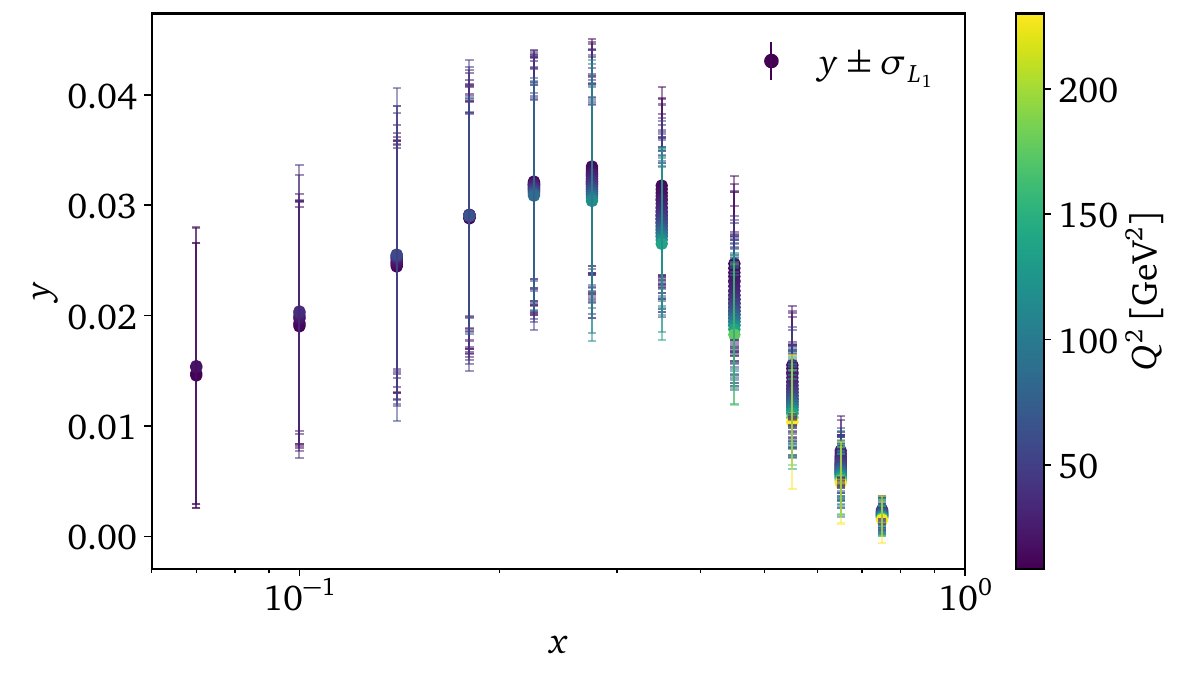}
    \caption{Illustration of $L_1$ data-space fluctuations. The central values show the underlying 
    noise-free theory prediction $y^\ast$, while the error bars indicate the standard deviation over 
    $1000$ independently generated $L_1$ replicas. The points are displayed as a function of their 
    $x$-grid position and coloured by $Q^2$.}
    \label{fig:y_data}
\end{figure}

As defined in Sec.~\ref{sec:t3model}, we either use noise-free $L_0$ data or one fluctuated $L_1$ closure-test replica 
to compare the different methods.
The $L_1$ data is obtained by adding a Gaussian fluctuation with covariance $C_{\rm exp}$ to the underlying theory 
prediction. Each random fluctation represents a possible realisation of the experimental data, accounting for the 
statistical and systematic uncertainties, namely a different run of the universe. 
Fig.~\ref{fig:y_data} illustrates the typical size of $L_1$ replica fluctuations by showing $1000$ independently 
generated $L_1$ replicas in data space.
The central points in Fig.~\ref{fig:y_data} correspond to the underlying noise-free theory prediction $y^\ast$, 
while the error bars indicate the standard deviation over the different $L_1$ replicas.
The points are shown as a function of the corresponding $x$-grid position and coloured according to their $Q^2$ value.
The spread is larger at small $x$, reflecting the larger range of possible pseudo-data values that the models may 
encounter in this kinematic region. The $1000$ replicas shown here are used only for illustration.
In the following comparisons, we use a single fixed $L_1$ realisation to compare all architectures on the same data, 
ensuring consistency and reproducibility of the results. 

First, we study how the choice of data-space loss affects NN reconstructions.
Thus, we start by considering the NN architecture trained with three different loss functions: a standard MSE loss, 
a heteroscedastic MSE loss (het MSE), and a covariance $\chi^2$ loss.
The MSE and heteroscedastic MSE losses are included as alternatives to the well-established $\chi^2$ loss. 
The plain MSE loss does not use any information on the data uncertainties, while the heteroscedastic MSE loss 
learns one variance $\sigma^2_{\theta,n}$ per data point, i.e.\ only the diagonal entries of a covariance matrix, 
without using the experimental uncertainties.
\begin{figure}[t!]
    \centering
    \includegraphics[width=0.49\textwidth, page=1]{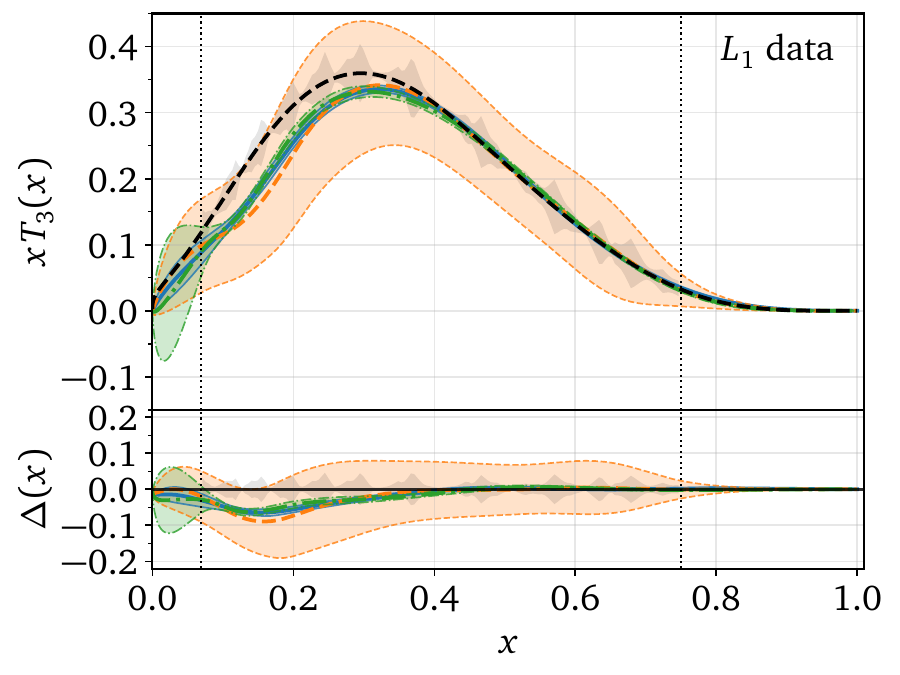}
    \includegraphics[width=0.49\textwidth, page=2]{figs/L1_NN.pdf}
    \caption{Comparing different loss functions for the NN on underlying $L_1$ data. Using a linear scale (left) and a log scale (right). The grey band is a regularised FK-projected visual reference for the approximate data-induced scale, which is not used as a fitted PDF uncertainty. The coloured bands show the $1\sigma$ intervals obtained from $C_\mathrm{ens}$. The vertical lines indicate the distinction between interpolation and extrapolation regions.}
    \label{fig:NN_loss}
\end{figure} 
Fig.~\ref{fig:NN_loss} shows the corresponding reconstructions of the underlying $T_3$ distribution from the fixed 
$L_1$ dataset.
The vertical dashed lines indicate the boundary between the interpolation region, where the FK map is directly 
constrained by the training data, and the extrapolation region, which is probed only at the prediction level.
The lower panels show the residual
\begin{align}
    \Delta(x)=f_{\rm pred}(x)-f(x) \eqcomma
    \label{eq:residual}
\end{align}
where $f_{\rm pred}(x)$ is the model prediction and $f(x)$ is the underlying closure-test truth.
The grey band shown in the PDF-space figures is included only as a visual diagnostic for the approximate scale 
of data-induced fluctuations after projection through the FK map as written in Eq.~\eqref{eq:discrete-forward-map}. 
It is not used as a fitted PDF uncertainty and should not be interpreted as a statistically complete uncertainty 
estimate. 
Since the inverse map from data space to PDF space is ill-conditioned (Sec.~\ref{subsec:FKT3}), we construct 
this visual reference with a regularised linear estimator. Defining, as in Sec.~\ref{subsec:FKT3}, 
\begin{align}
    A = W^{T} C_{\rm exp}^{-1} W \eqcomma
\end{align}
with $C_{\rm exp}$ the experimental covariance matrix in data space and $W$ the corresponding FK-table matrix. 
The matrix $A$ coincides with the matrix $M$ of Ref.~\cite{Chiefa:2025cap} and is singular or severely ill-conditioned. 
Its inversion therefore requires either a Tikhonov regularisation, as adopted here, or a pseudo-inverse restricted to 
the image of $A$, which requires a numerical tolerance separating ${\rm Im}(A)$ from ${\rm Ker}(A)$~\cite{Chiefa:2025cap}.
We therefore introduce
\begin{align}
    \hat f_{\rm lin} = (A+\lambda \mathbb{I})^{-1} W^T C_{\rm exp}^{-1} \mathbf{y} \eqcomma
\end{align}
where $\lambda > 0$ controls the regularisation strength.
For a closure-test fluctuation 
$\mathbf{y}=Wf^\ast+\boldsymbol{\eta}$, with $\boldsymbol{\eta}\sim\mathcal{N}(0,C_{\rm exp})$, the covariance of 
the noise-induced fluctuation propagated through this regularised linear map is
\begin{align}
    C_{\rm W}^{(\lambda)} = (A+\lambda \mathbb{I})^{-1} A (A+\lambda \mathbb{I})^{-1} \eqperiod
\end{align}
We use this quantity only to draw the grey reference band. 
The regularisation parameter is set to
\begin{align}
    \lambda = 0.1\, \frac{\mathrm{Tr}(A)}{N_{\rm grid}} \eqcomma
\end{align}
which fixes the ridge scale to ten percent of the average eigenvalue scale of $A$. The conclusions below do not 
rely on the precise normalisation of this visual band.
The difference between the three loss functions is also visible directly in the observable space, where the loss 
is evaluated.

\begin{figure}[t!]
    \centering
    \includegraphics[width=0.49\textwidth]{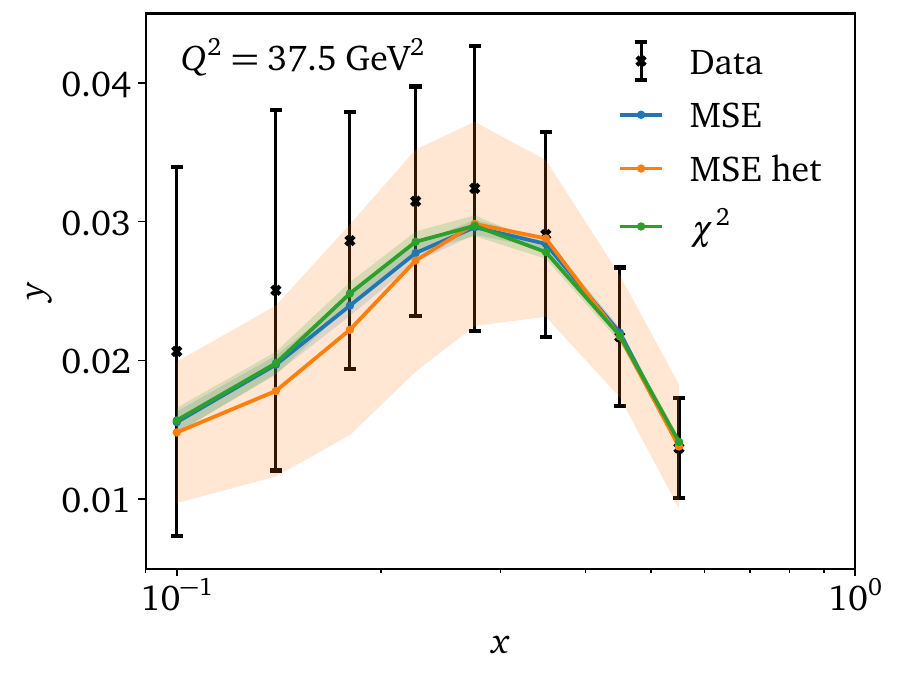}
    \caption{Observable space comparison for a fixed $Q^2$ slice. The points indicate the reference 
    data space values, while the model predictions are shown for the MSE, heteroscedastic MSE, and $\chi^2$-trained NN setups. The displayed uncertainties combine the ensemble spread with the heteroscedastic contribution where applicable.}
    \label{fig:NN_ypred}
\end{figure}

Fig.~\ref{fig:NN_ypred} compares the model predictions for a fixed $Q^2$ slice with the corresponding data space 
reference values. The three different loss functions all provide a similarly stable description of the 
underlying $L_1$-data.
\begin{figure}[b!]
    \centering
    \includegraphics[width=0.49\textwidth, page=1]{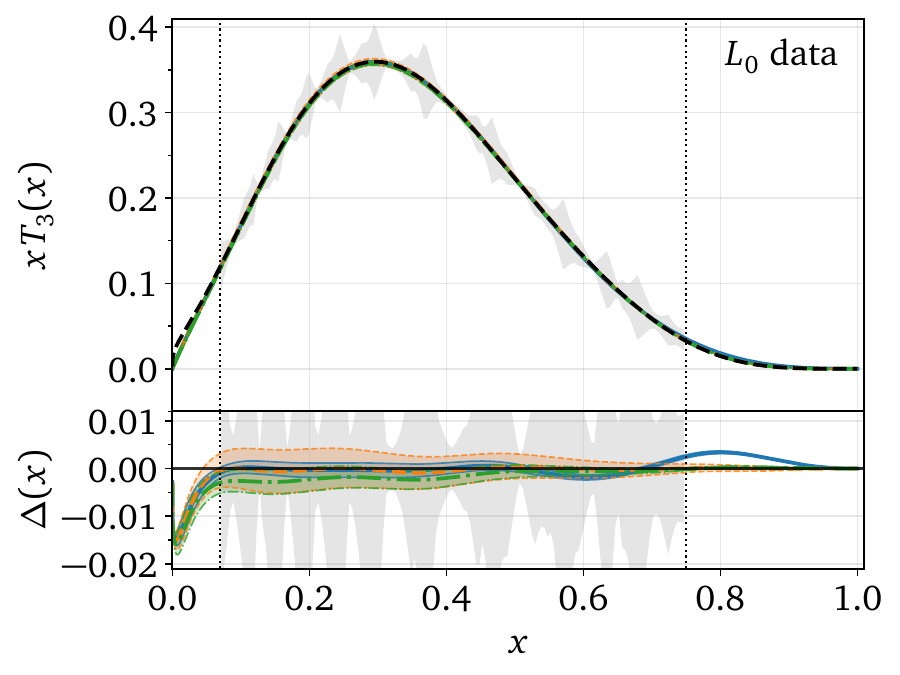}
    \includegraphics[width=0.49\textwidth, page=2]{figs/L0_NN.pdf}
    \caption{Comparison of NN loss functions trained on noise-free $L_0$ data, shown on a linear $x$-scale (left) and logarithmic $x$-scale (right). The vertical lines indicate the distinction between interpolation and extrapolation regions.}
    \label{fig:NN_L0}
\end{figure} 
For the heteroscedastic MSE, the learned pointwise variance $\sigma^2_{\theta,n}$ absorbs the fluctuations of 
the $L_1$ data, and the resulting uncertainty band is large in both the interpolation and the extrapolation regions.
Its large uncertainty bands should therefore be interpreted as a symptom of likelihood misspecification rather than 
as a faithful PDF uncertainty estimate.
The $\chi^2$-trained NN, on the other hand, provides a smaller uncertainty in the interpolation region and inflates 
the model uncertainty in the extrapolation region, while the MSE-trained NN yields the smallest uncertainty over 
the whole range of $x$, as expected for a loss that does not use any information on the data uncertainties.

For comparison, Fig.~\ref{fig:NN_L0} shows the same NN loss comparison when the networks are trained on noise-free 
$L_0$ data.
In this case, the differences between the loss functions are reduced because the target data no longer contains a 
random $L_1$ fluctuation.
All three losses give nearly indistinguishable results and recover the truth in the interpolation region. 
In the extrapolation region, they share a common deviation from the truth: $xT_3$ vanishes faster than the 
truth as $x\to0$, and the uncertainty bands do not cover it. 
Since the data are noise-free, this deviation cannot be attributed to the loss function, but rather to the 
small-$x$ preprocessing, 
since $\alpha\simeq1$ forces $xT_3\propto x$ at small $x$ (see Fig.~\ref{fig:alpha_scan_fixed_beta}).

\begin{figure}[t!]
    \centering
    \includegraphics[width=0.49\textwidth, page=1]{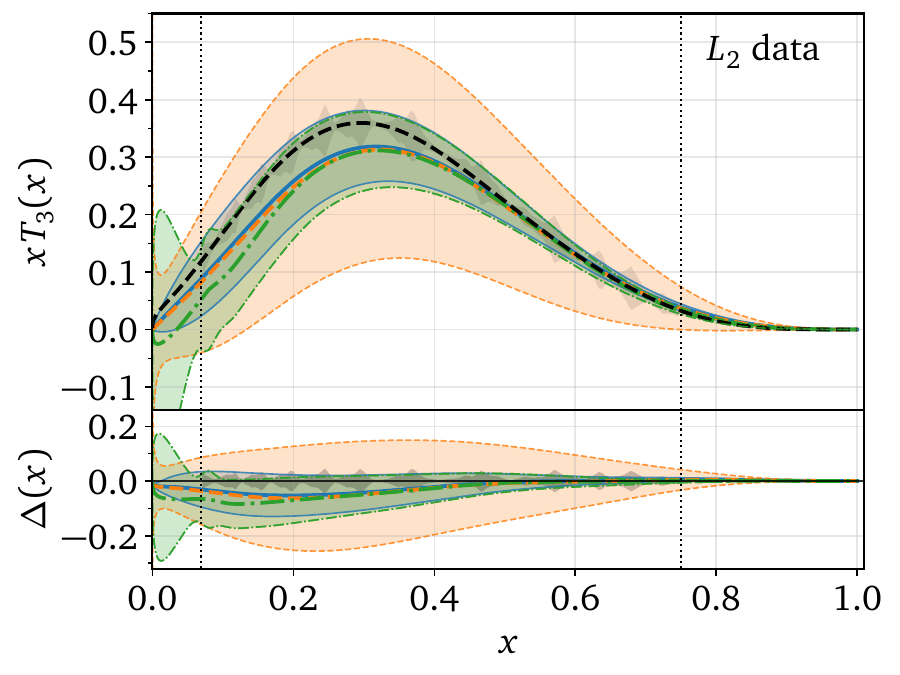}
    \includegraphics[width=0.49\textwidth, page=2]{figs/L2_NN_comp.pdf}
    \caption{Comparison of NN loss functions trained on $L_2$ data, shown on a linear $x$-scale (left) and logarithmic $x$-scale (right).}
    \label{fig:NN_L2}
\end{figure}

Fig.~\ref{fig:NN_L2} shows the same comparison for NNs trained on $L_2$ replicas. 
In the interpolation region, the MSE and $\chi^2$ losses give similar uncertainty bands which, unlike in 
the $L_1$ case, cover the truth around the peak of the distribution, 
as expected since the $L_2$ spread includes the propagated data uncertainty (Sec.~\ref{sec:unc_quant}). 
In the small-$x$ extrapolation region, the $\chi^2$ uncertainty grows substantially, while the MSE band remains narrow. 
The heteroscedastic MSE again yields a very large uncertainty over the whole range of $x$.
\begin{figure}[b!]
    \centering
    \includegraphics[width=0.49\textwidth, page=1]{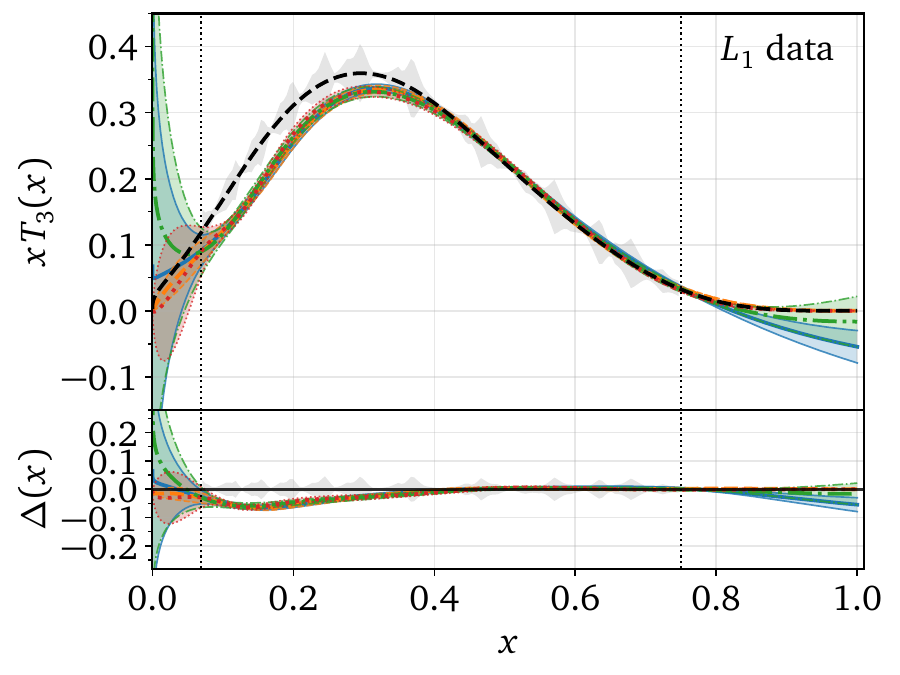}
    \includegraphics[width=0.49\textwidth, page=2]{figs/NN_L1_arc.pdf}
    \caption{Effect of endpoint preprocessing for NN fits to $L_1$ data. The ``raw'' setups use the baseline NN, while the ``full'' setups include the endpoint prefactor $x^\alpha(1-x)^\beta$. Results are shown for MSE and $\chi^2$ training on a linear $x$-scale (left) and logarithmic $x$-scale (right).}
    \label{fig:NN_L1_arc}
\end{figure} 
Next, we assess the impact of the trainable endpoint prefactor introduced in Sec.~\ref{sec:setup}.
In Fig.~\ref{fig:NN_L1_arc}, the setups labelled ``raw'' use the plain NN with the baseline input preprocessing, 
while the setups labelled ``full'' include the multiplicative endpoint factor $x^\alpha(1-x)^\beta$, with $\alpha$ 
and $\beta$ treated as trainable parameters within the ranges specified in Sec.~\ref{sec:setup}.
The comparison is performed for both the MSE and $\chi^2$ losses on the same $L_1$ dataset.
\begin{figure}[t!]
    \centering
    \includegraphics[width=0.49\textwidth, page=1]{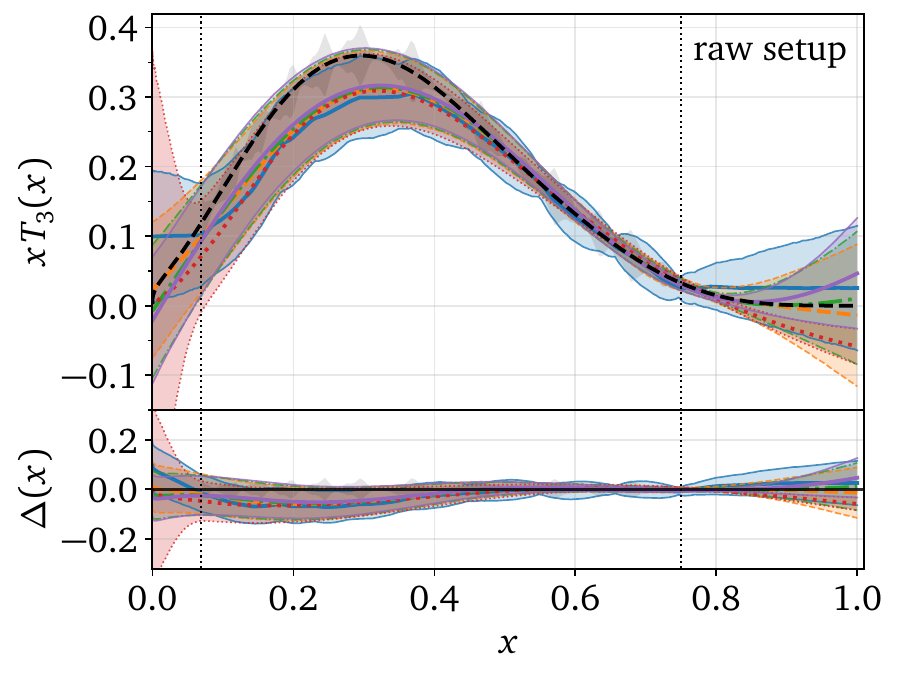}
    \includegraphics[width=0.49\textwidth, page=2]{figs/GP_raw.pdf}
    \caption{Raw GP reconstructions using different kernel choices for $L_1$ data. The raw setup does not include the endpoint prefactor. Results are shown on a linear $x$-scale (left) and logarithmic $x$-scale (right).}
    \label{fig:GP_raw}
\end{figure} 
\begin{figure}[b!]
    \centering
    \includegraphics[width=0.49\textwidth, page=1]{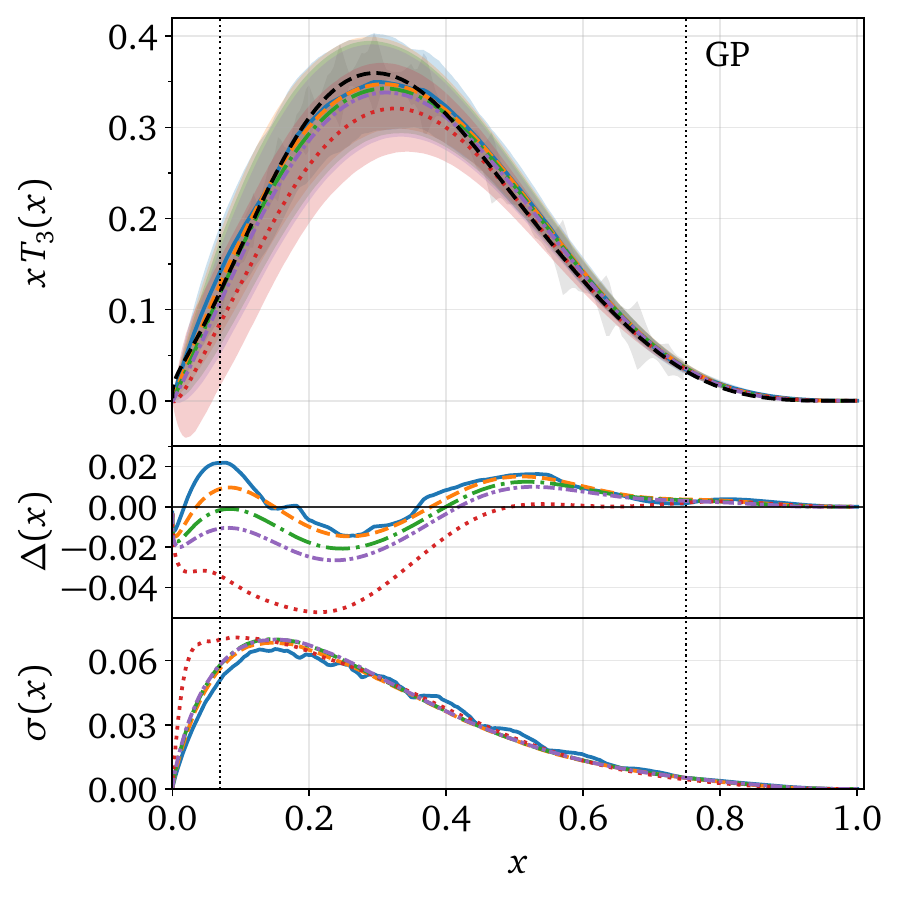}
    \includegraphics[width=0.49\textwidth, page=2]{figs/GP_full2.pdf}
    \caption{GP reconstruction with endpoint preprocessing included using $L_1$. Results are shown for different kernels on a linear $x$-scale (left) and a logarithmic $x$-scale (right). Below the $xT_3(x)$ prediction, the residual $\Delta(x)$ and the $1\sigma$ predictive uncertainty are shown.}
    \label{fig:GP_full}
\end{figure} 

For both networks, the endpoint prefactor affects only the description of the extrapolation region.
For the MSE-trained network, its main effect is to stabilise the transition between interpolation and 
extrapolation for large $x$ values.
This indicates that even a simple physics-inspired endpoint structure can act as a useful regulariser of 
the inverse problem in regions where the data do not strongly constrain the PDF. 
At the same time, the extrapolation obtained with the prefactor is largely determined by the assumed exponents: 
as shown in Fig.~\ref{fig:alpha_scan_fixed_beta}, 
changing the small-$x$ exponent from $\alpha=1.0$ to $\alpha=0.5$ at fixed $\beta$ leaves the interpolation 
region essentially unchanged, 
but substantially modifies both the central value and the uncertainty below $x\simeq0.07$. 
The small-$x$ uncertainty bands should therefore be understood as conditional on the preprocessing.

Next, we consider GPs with different kernel choices.
In Fig.~\ref{fig:GP_raw} we show the raw GP setup for several kernel choices: RBF, Gibbs, and Mat\'ern kernels 
with different values of the smoothness parameter $\nu$.
In this raw configuration, the GP prior is not augmented by the endpoint prefactor.
The resulting central predictions do not fully reconstruct the peak region from the underlying $L_1$ data, and 
the extrapolation region shows sizeable kernel-dependent uncertainties.
This demonstrates that in the present inverse problem, the choice of kernel alone is insufficient to encode 
the relevant physical structure.

In Fig.~\ref{fig:GP_full} we display the corresponding GP results after including the same endpoint preprocessing 
used in the full NN setup.
The endpoint preprocessing substantially improves peak-region reconstruction and reduces residuals in the interpolation 
region.
The behaviour in the extrapolation region is also more stable than in the raw setup.
The lower two panels show the residual $\Delta(x)$ of Eq.~\eqref{eq:residual} and the $1\sigma$ predictive uncertainty 
$\sigma(x)$. 
The uncertainty vanishes at both endpoints. 
This is a direct consequence of the preprocessing: since $k_f(x,x)=x^{2\alpha}(1-x)^{2\beta}\,k_h(x,x)$, the prior 
variance of $xT_3$, and hence its posterior variance, 
is suppressed as $x\to0$ and $x\to1$ by construction. 
At small $x$ this results in uncertainty bands that do not cover the truth, which, as for the NNs trained on 
$L_0$ data, deviates from the reconstruction because of the assumed small-$x$ exponent. 
The narrow small-$x$ bands of the preprocessed GPs should therefore not be interpreted as a data-driven precision.

\begin{figure}[t!]
    \centering
    \includegraphics[width=0.49\textwidth, page=1]{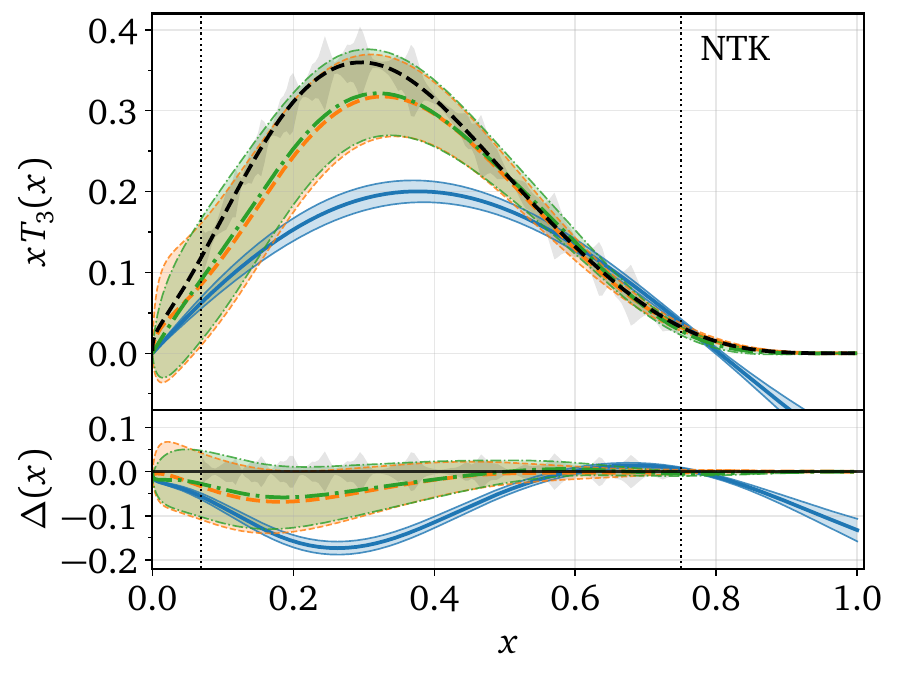}
    \includegraphics[width=0.49\textwidth, page=2]{figs/ntk_test.pdf}
    \caption{Empirical finite-width NTK-based reconstructions evaluated at random initialisation, after 100 training epochs, and after the full 5000-epoch training run. All three NTKs are computed from the same NN architecture and $\chi^2$-training setup. We evaluate the initialisation NTK before optimisation, whereas the other two are evaluated along their respective training trajectories.}
    \label{fig:NTK_full}
\end{figure} 
As an interface between NNs and kernel methods, we also study NTK-based reconstructions.
In this work, the NTK should be understood as an empirical finite-width kernel induced by the NN architecture, 
rather than as a closed-form infinite-width kernel.
We compare three linearised NTK-based predictions in Fig.~\ref{fig:NTK_full}, one obtained by linearising around 
the random initialisation, one obtained by linearising around the parameters after 100 training epochs, and one 
obtained by linearising around the parameters after the full 5000-epoch training run.
All three empirical NTKs are computed from the same NN architecture and training setup used for the full 
$\chi^2$-trained NN with the hyperparameters listed in App.~\ref{app:hyper}.
The first is evaluated before optimisation, while the latter two are evaluated along their respective training 
trajectories.
The initialisation NTK significantly underestimates the true distribution and does not reproduce the peak structure. 
This shows that, for the finite-width architecture and hyperparameters used here, the initialisation NTK is not a 
faithful surrogate for the trained nonlinear network.
Evaluating the empirical NTK after a small number of training epochs already improves the central prediction, 
indicating that the local geometry changes rapidly during early training. 
After the full training run, the NTK prediction only slightly improves in the interpolation region and yields a 
narrower uncertainty estimate in the extrapolation region compared to the NTK evaluation after a shorter training 
period.
This suggests that the trained empirical kernel captures part, but not all, of the behaviour of the nonlinear 
trained network.

\begin{figure}[t!]
    \centering
    \includegraphics[width=0.49\textwidth, page=1]{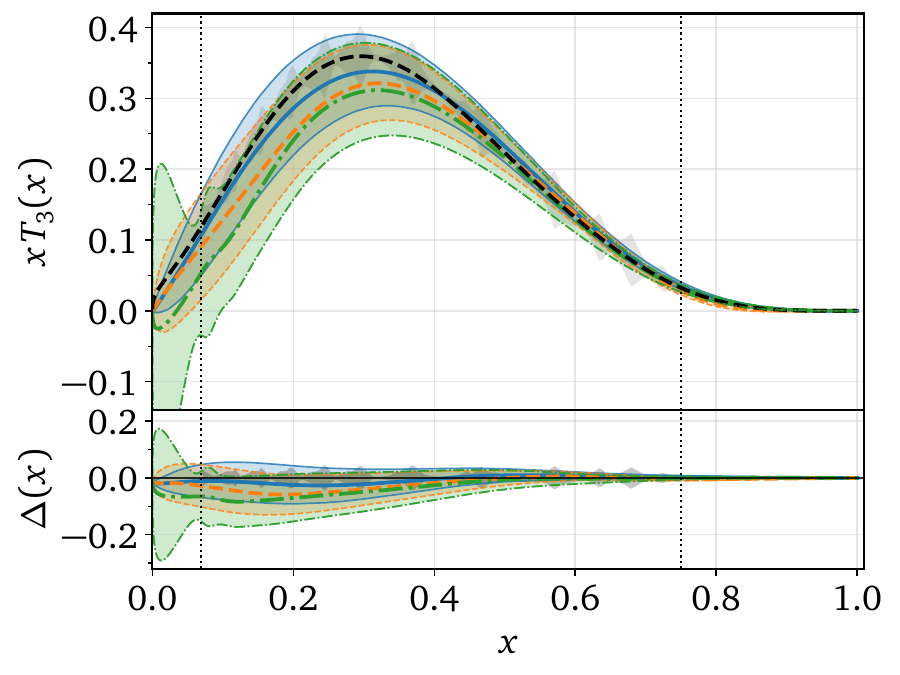}
    \includegraphics[width=0.49\textwidth, page=2]{figs/all_L2_new.pdf}
    \caption{Representative comparison of NN, GP, and NTK reconstructions. The NTK and GP results are trained or conditioned on $L_1$ data, while the additional $\chi^2$-NN $L_2$ result uses $L_2$ replicas as a simplified analogue of an NNPDF-style replica fit.}
    \label{fig:model_comp}
\end{figure}

Finally, Fig.~\ref{fig:model_comp} compares representative setups from each model class.
For the GP, we show the RBF kernel, which gives one of the most stable reconstructions among the kernels considered.
For the NTK, we use the fully trained NTK, since using only the initialised NTK fails to reproduce the central structure.
The $L_2$-trained $\chi^2$ NN is included as a central-value comparison to a simplified replica-based strategy. 
Since the $L_2$ setup is a simplified analogue of the NNPDF methodology (Sec.~\ref{sec:NN}), its ensemble covariance 
should not be interpreted as a full NNPDF-style PDF uncertainty.
The trained NTK, GP, and $L_2$-trained $\chi^2$ NN all recover the underlying truth within their respective 
uncertainty estimate in the interpolation region.
Among the representative setups shown, the RBF GP yields the closest central reconstruction of the underlying truth.
However, the RBF GP uncertainty band becomes very narrow near the endpoints because of the preprocessing discussed 
above, while the $L_2$ $\chi^2$-trained NN uncertainty grows in the extrapolation region.
This again illustrates that an accurate central prediction does not by itself imply a faithful uncertainty estimate, 
especially outside the data-supported region. 

\subsection{Faithfulness of uncertainties in the data region}
\label{sec:bias_res}
We start by testing the faithfulness of the uncertainty estimates in the data region, $x\gtrsim0.07$, where the 
models are trained and the reconstructions are directly constrained by the BCDMS data.
To this purpose, we use the normalised-bias diagnostic defined in Eq.~\eqref{eq:normalised-bias},
evaluated in the interpolation region with either the full ensemble covariance $C_{\rm ens}$ or its diagonal 
$\mathrm{diag}(C_{\rm ens})$. 
In both cases, we regularise the covariance by retaining the leading eigenvectors that account for at least 
$99.9\%$ of its trace, and we normalise the diagnostic by the number $n_{\rm eig}$ of retained eigenvalues.
The corresponding results are shown in Tab.~\ref{tab:normalised-bias}.
For the full covariance, $n_{\rm eig}$ ranges between 4 and 19 depending on the model, so that $R_b$ probes 
only a small number of directions in data space 
and, as discussed in Sec.~\ref{sec:unc_quant}, its statistical precision on a single realisation is limited.

\begin{table}[t!]
    \centering
    \begin{small} \begin{tabular}{l|cc|cc}
        \toprule
         model & $R_b$, $C_{\rm ens}$ & $R_b$, $\mathrm{diag}(C_{\rm ens})$ & $n_{\rm eig}$, $C_{\rm ens}$ & $\sigma_{R_b}$ \\ 
         \midrule
         NNs ($L_1$)          & 2.844 & 3.131 & 11 & 0.213 \\
         Gibbs                & \textbf{0.856} & 0.529 & 6  & 0.289 \\
         RBF                  & \textbf{1.083} & 0.384 & 4  & 0.354 \\
         Matern $(\nu = 0.5)$ & 0.625 & 0.472 & 14 & 0.189 \\
         Matern $(\nu = 1.5)$ & \textbf{0.922} & 0.445 & 7  & 0.267 \\
         Matern $(\nu = 2.5)$ & \textbf{1.154} & 0.406 & 5  & 0.316 \\
         NTK init             & 9.505 & 7.860 & 4  & 0.354 \\
         NTK trained          & 0.674 & 0.669 & 19 & 0.162 \\
         \midrule
         NNs ($L_2$)          & \textbf{0.743} & 0.587 & 7 & 0.267 \\
         \bottomrule
    \end{tabular} \end{small}
    \caption{Normalised bias $R_b$, Eq.~\eqref{eq:normalised-bias}, for different models, evaluated using either the full ensemble covariance $C_{\rm ens}$ or its diagonal approximation $\mathrm{diag}(C_{\rm ens})$, regularised by retaining the $n_{\rm eig}$ leading eigenvectors that account for $99.9\%$ of the trace. The last two columns report the number of retained eigenvalues for the full covariance and the corresponding approximate fluctuation scale, $\sigma_{R_b}=1/\sqrt{2n_{\mathrm{eig}}}$. Values in bold are compatible with a faithful uncertainty at the 95\% confidence level, given $n_{\rm eig}$. The diagnostic is evaluated only in the interpolation region. The trained NTK is evaluated after 5000 epochs, while the initialisation NTK is evaluated before training.}
    \label{tab:normalised-bias}
\end{table}
The comparison demonstrates that covariance correlations can have a substantial and model-dependent effect on 
the diagnostic.
These values should be read in light of the cumulative interpretation of the closure-test levels discussed in 
Sec.~\ref{sec:unc_quant}. 
The $L_1$ ensemble is trained on a single fixed data realisation with different initialisations, so that its 
spread contains the functional uncertainty 
but not the propagated data uncertainty. 
The deviation of its mean from the truth, which is dominated by the $L_1$ fluctuation $\boldsymbol{\eta}$, 
is therefore not expected to be covered by $C_{\rm ens}$. 
When the data uncertainty is included, either through the $L_2$ replicas or, for the GPs and the trained NTK, 
through the experimental covariance in the likelihood, 
$R_b$ is of order one, as discussed below.
Across the models, however, neglecting correlations does not lead to a systematic increase or decrease in $R_b$. 
Its effect depends on the alignment between the residuals and the retained covariance directions.

The GP results exhibit the same model dependence.
The Gibbs kernel gives $R_b=0.856$ with the full covariance, close to unity, while the diagonal approximation 
reduces the value to $0.529$.
For the RBF kernel, both prescriptions yield $R_b=1.083$ and $0.384$, respectively.
The Mat\'ern kernels likewise give larger values with the full covariance than with the diagonal approximation.
As the smoothness parameter increases from $\nu=0.5$ to $\nu=2.5$, the full-covariance result increases from 
$0.625$ to $1.154$, with $R_b=0.922$ for $\nu=1.5$.
The corresponding diagonal results decrease moderately from $0.472$ to $0.406$.
All GP results with the full covariance are thus compatible with a faithful uncertainty at the 95\% confidence 
level, given the corresponding values of $n_{\rm eig}$, except for the Mat\'ern kernel with $\nu=0.5$, whose 
value $R_b=0.625$ lies marginally below this range, indicating a slightly conservative uncertainty.

The largest normalised bias occurs at initialisation for the NTK, with $R_b=9.505$ using the full covariance 
and $R_b=7.860$ using its diagonal.
Training substantially improves these values, reducing them to $0.674$ and $0.669$, respectively.
The NTK at initialisation is the only setup that includes the data uncertainty and nevertheless underestimates 
it substantially, 
since its kernel and prior mean are evaluated at random parameter values that are poorly adapted to the problem, 
whereas the trained NTK gives $R_b=0.674$, marginally below the 95\% range for $n_{\rm eig}=19$, indicating a 
slightly conservative uncertainty.

Overall, none of the setups that include the data uncertainty underestimates it, apart from the NTK at 
initialisation: 
their values of $R_b$ are compatible with a faithful uncertainty at the 95\% confidence level or, for the Mat\'ern 
kernel with $\nu=0.5$ and the trained NTK, 
marginally below this range, indicating slightly conservative uncertainties. 
In contrast, the $L_1$ ensemble, which by construction excludes data uncertainty, and the NTK at initialisation 
give values far above one. 
Since $n_{\rm eig}$ is small, this test is not very constraining: it excludes a substantial underestimate of the 
uncertainty, but cannot establish its faithfulness.
The changes observed when replacing $C_{\rm ens}$ with $\mathrm{diag}(C_{\rm ens})$ show that calibration depends 
not only on the marginal variances 
but also on the correlation structure of the ensemble predictions. 
These conclusions apply to the retained subspace in the interpolation region and to the present realisation only; 
in the extrapolation region, as shown in Sec.~\ref{sec:arc_res}, the uncertainty is conditional on the preprocessing.
Finally, all results in this section refer to a single $L_1$ data realisation, generated with a fixed random seed. 
As shown in App.~\ref{app:L2} (Fig.~\ref{fig:NN_diffseed}), a different seed visibly changes the reconstruction, 
in particular without endpoint preprocessing, 
so that the values of $R_b$ in Tab.~\ref{tab:normalised-bias} are subject to realisation-to-realisation fluctuations. 

If the test were repeated on $N_{L_1}$ independent $L_1$ realisations, i.e.\ on different ``runs of the universe'', 
the statistical uncertainty on the average of $R_b$
would decrease as $1/\sqrt{N_{L_1}}$, from $1/\sqrt{2n_{\rm eig}}$ to approximately $1/\sqrt{2n_{\rm eig}N_{L_1}}$,
so that the diagnostic would become sensitive to much smaller miscalibrations than those that can be resolved here.
A systematic study over many $L_1$ realisations, which would determine the distribution of $R_b$, is left for future 
work.
\subsection{Faithfulness of uncertainties in the extrapolation region}
\label{sec:hera_res}
The normalised bias of Sec.~\ref{sec:bias_res} tests the uncertainties only where the models are trained.
As a final test, we probe the reconstructed $T_3$ and its uncertainty in the extrapolation region by inserting 
it into HERA reduced-cross-section predictions.
HERA measurements extend to values of $x$ well below the range covered by BCDMS, $x\gtrsim0.07$, so that a sizeable 
fraction of the HERA data lies in a region of $x$
that was not used to train the models.
This comparison therefore tests the extrapolation of the reconstructed $T_3$ and of its uncertainty, rather 
than the region directly constrained by the training data.
We consider two neutral-current HERA datasets: the electron-beam NC EM dataset at $\sqrt{s}=318\,{\rm GeV}$ and 
the positron-beam NC EP dataset at $\sqrt{s}=251\,{\rm GeV}$.

The HERA reduced cross section contains several PDF contributions. We write the theory prediction schematically as
\begin{align}
    T_{\rm HERA} = T_{3,{\rm HERA}} + R_{\rm HERA} \eqcomma
    \label{eq:hera_pred}
\end{align}
where $T_{3,{\rm HERA}}$ denotes the native triplet contribution and $R_{\rm HERA}$ collects all remaining 
PDF contributions.
The replacement test keeps $R_{\rm HERA}$ fixed and substitutes only the triplet component,
\begin{align}
    T_{\rm pred} = T_{3,{\rm model}} + R_{\rm HERA} \eqcomma
\end{align}
where $T_{3, \mathrm{model}}$ is obtained from the BCDMS-trained NN, GP or NTK reconstruction. 
For the NNs we use the ensembles trained on $L_2$ replicas, whose spread includes the data uncertainty 
(Sec.~\ref{sec:unc_quant}), 
while the GP and NTK reconstructions are conditioned on the $L_1$ data.
This replacement test is intentionally restricted, with all non-$T_3$ PDF contributions held fixed. 
It therefore does not account for compensating shifts or correlations among PDF flavours that would arise 
in a simultaneous global fit. 
The test should be viewed as a targeted propagation and compatibility check for the reconstructed $T_3$ component, 
not as a replacement for a full refit.
The agreement with the HERA data is quantified using the reduced $\chi^2$ function
\begin{align}
    \chi^2_{\rm red}=\frac{1}{N_{\rm data}^{\rm HERA}}(\mathbf{y}_{\rm HERA}-G)^T C^{-1}(\mathbf{y}_{\rm HERA}-G) \eqcomma
\end{align}
where $\mathbf{y}_{\rm HERA}$ denotes HERA Level-1 pseudo-data, generated from the closure-test truth with the 
experimental covariance $C^{\rm HERA}_{\rm exp}$, 
$G$ is either the native or modified theory prediction, and $C$ is the covariance used in the comparison, 
either $C^{\rm HERA}_{\rm exp}$ or $C^{\rm HERA}_{\rm tot}$ defined below.
The baseline prediction is computed with the NNPDF4.0 PDFs, i.e.\ with the same underlying truth.
When model-spread uncertainties are included, we use
\begin{align}
    C_{\rm tot}^{\rm HERA} = C_{\rm exp}^{\rm HERA} + C_{\rm ens}^{\rm HERA} \eqperiod
\end{align}
Here $C_{\rm ens}^{\rm HERA}$ is obtained by propagating the model ensemble through the HERA FK map for the 
replaced $T_3$ contribution only, while the remaining contribution $R_{\rm HERA}$ is held fixed.
For a comparison of the native and replaced HERA contributions, we define
\begin{align}
    \Delta \chi^2_{\rm red}=\chi^2_{{\rm red},\,{\rm pred}} - \chi^2_{{\rm red},\,{\rm base}} \eqcomma
\end{align}
where the baseline corresponds to the unmodified HERA theory prediction.
\begin{table}[b!]
    \centering
    \begin{small}
            \renewcommand{\arraystretch}{1.25}
    \vspace{0.5em}
    \begin{tabular}{ll|cc}
        \toprule
         Prediction NC EM & $C$ & $\Delta \chi^2_{\rm red}$ & $\Delta n_\sigma$\\
         \midrule
         NN $\chi^2$-loss & $C_{\rm exp}^{\rm HERA}$
         & 0.101 & 0.899 \\
         NN $\chi^2$-loss & $C_{\rm tot}^{\rm HERA}$
         & -0.008 & -0.075 \\
         NN MSE-loss & $C_{\rm exp}^{\rm HERA}$
         & 0.051 & 0.457 \\
         NN MSE-loss & $C_{\rm tot}^{\rm HERA}$
         & -0.017 & -0.153 \\
         GP RBF kernel & $C_{\rm exp}^{\rm HERA}$
         & -0.002 & -0.018 \\
         GP RBF kernel & $C_{\rm tot}^{\rm HERA}$
         & -0.013 & -0.114 \\
         NTK at init & $C_{\rm exp}^{\rm HERA}$
         & 0.558 & 4.973 \\
         NTK at init & $C_{\rm tot}^{\rm HERA}$
         & 0.272 & 2.424 \\
         NTK after training & $C_{\rm exp}^{\rm HERA}$
         & 0.060 & 0.535 \\
         NTK after training & $C_{\rm tot}^{\rm HERA} $
         & -0.055 & -0.495 \\
         \midrule
         Baseline & $C_{\rm exp}^{\rm HERA}$ & 1.092 & $N_\mathrm{data} = 159$ \\
         \bottomrule
    \end{tabular}
    \vspace{1.5em}
    \vspace{1.5em}
    \begin{tabular}{ll|cc}
    \toprule
         Prediction NC EP & $C$ & $\Delta \chi^2_{\rm red}$ & $\Delta n_\sigma$\\
         \midrule
         NN $\chi^2$-loss & $C_{\rm exp}^{\rm HERA}$
         & 0.005 & 0.060 \\
         NN $\chi^2$-loss & $C_{\rm tot}^{\rm HERA}$
         & -0.0001 & -0.001 \\
         NN MSE-loss & $C_{\rm exp}^{\rm HERA}$
         & 0.003 & 0.030 \\
         NN MSE-loss & $C_{\rm tot}^{\rm HERA}$
         & 0.0002 & 0.003 \\
         GP RBF kernel & $C_{\rm exp}^{\rm HERA}$
         & -0.001 & -0.006 \\
         GP RBF kernel & $C_{\rm tot}^{\rm HERA}$
         & -0.001 & -0.013 \\
         NTK at init & $C_{\rm exp}^{\rm HERA}$
         & 0.020 & 0.224 \\
         NTK at init & $C_{\rm tot}^{\rm HERA}$
         & 0.019 & 0.213 \\
         NTK after training & $C_{\rm exp}^{\rm HERA}$
         & 0.003 & 0.039 \\
         NTK after training & $C_{\rm tot}^{\rm HERA}$
         & -0.0001 & -0.002 \\
         \midrule
         Baseline & $C_{\rm exp}^{\rm HERA}$ & 0.912 & $N_\mathrm{data} = 254$ \\
         \bottomrule
        \end{tabular}

    \end{small}

    \caption{Shift in $\Delta \chi^2_{\rm red}$ and corresponding $\Delta n_\sigma$ for the HERA NC EM dataset at $\sqrt{s}=318\,{\rm GeV}$ (left) and the HERA NC EP dataset at $\sqrt{s}=251\,{\rm GeV}$ (right). The baseline is the unmodified HERA theory prediction, while the substituted predictions replace only the $T_3$ contribution by the corresponding BCDMS-trained model prediction. The NN results are obtained from ensembles trained on $L_2$ replicas.}
    \label{tab:hera_f3_nc_em_ep}
\end{table}
We also quote the shift in units of the expected standard deviation of a reduced $\chi^2$,
\begin{align}
    \Delta n_\sigma = \frac{\Delta \chi^2_{\rm red}}{\sqrt{2/N_{\rm data}^{\rm HERA}}} \eqperiod
\end{align}
Since the baseline coincides with the underlying truth, any deviation of the reconstructed $T_3$ from the truth 
increases the $\chi^2$ on average, 
so that the expected value of $\Delta\chi^2_{\rm red}$ is non-negative when the experimental covariance is used for both predictions, assuming that the BCDMS reconstruction is independent of the HERA noise. 
Small negative values can arise from the specific Level-1 fluctuation of the HERA pseudo-data, while with $C^{\rm HERA}_{\rm tot}$ the shifts are further reduced by the enlarged covariance entering the $\chi^2$. 
The resulting shifts are shown in Tab.~\ref{tab:hera_f3_nc_em_ep} for the NC EM (left) and NC EP (right) datasets.

For NC EM, the shifts have both signs. With the experimental covariance, the NN $\chi^2$, NN MSE and trained NTK 
reconstructions give $\Delta n_\sigma=0.899$, $0.457$ and $0.535$, respectively, while the RBF GP gives a small 
negative shift, $-0.018$.
The NTK at initialisation produces the largest shift, $\Delta n_\sigma=4.973$.
Including the propagated model covariance makes the shifts negative for all other models, ranging from $-0.495$ 
to $-0.075$, and reduces the initialisation-NTK shift to $2.424$. 
For NC EP, all shifts remain below $0.25$ in absolute value.
The NTK at initialisation thus produces the largest deterioration in NC EM agreement under both covariance 
prescriptions, consistent with its large normalised bias 
in the interpolation region, while the remaining models produce shifts smaller than one unit of the reference 
fluctuation scale in both datasets.
The native predictions for both datasets give $\chi^2_{\mathrm{red}}=1.092$ for the NC EM dataset and $0.912$ 
for the NC EP dataset, both within one standard deviation of the reference fluctuation 
scale $\sqrt{2/N_{\mathrm{data}}}$.

\begin{figure}[t!]
    \centering
    \includegraphics[width=0.49\textwidth]{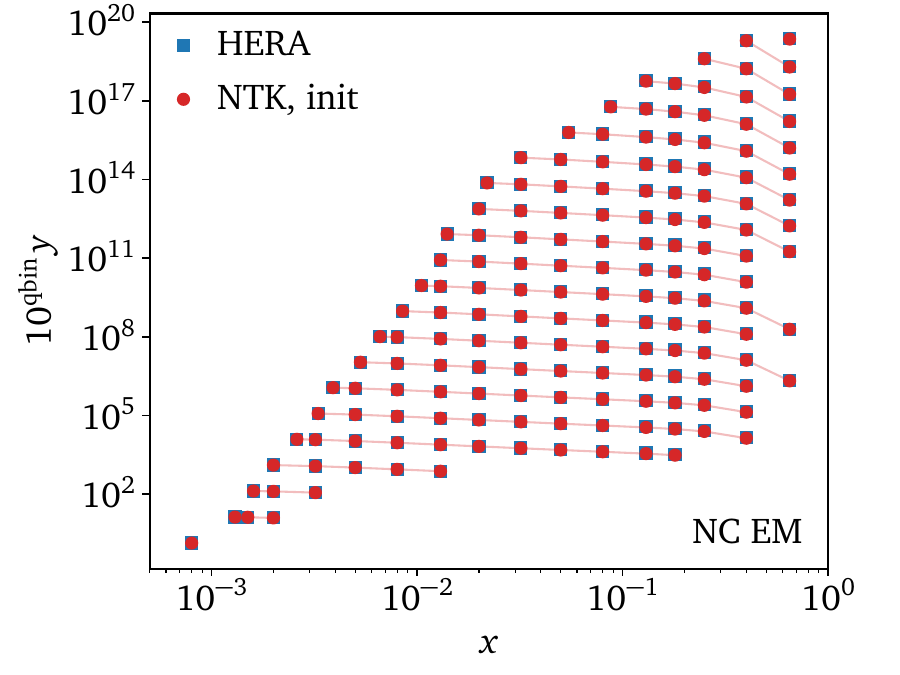}
    \includegraphics[width=0.49\textwidth]{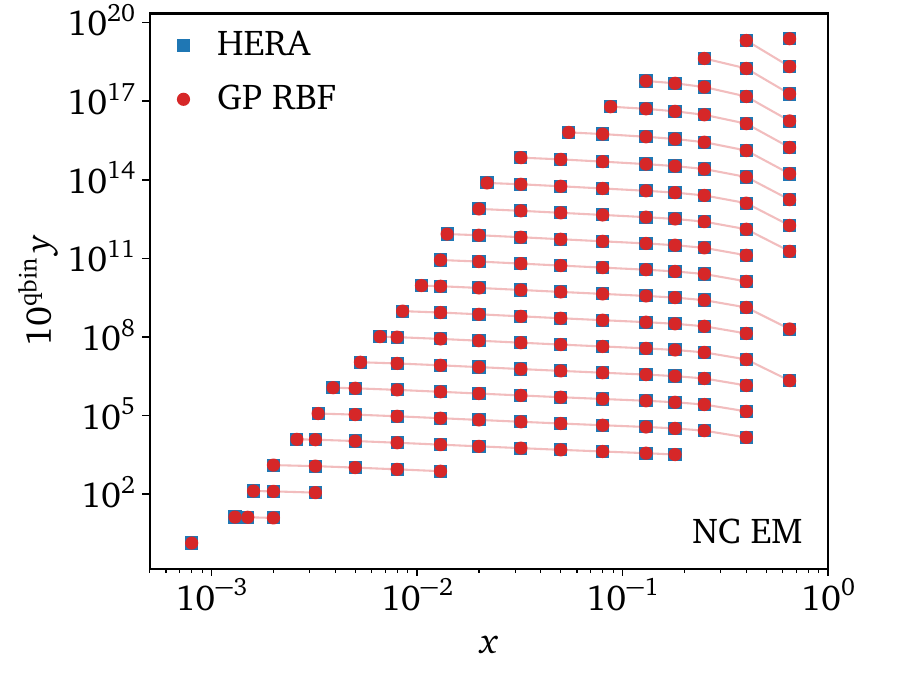}
    \includegraphics[width=0.49\textwidth]{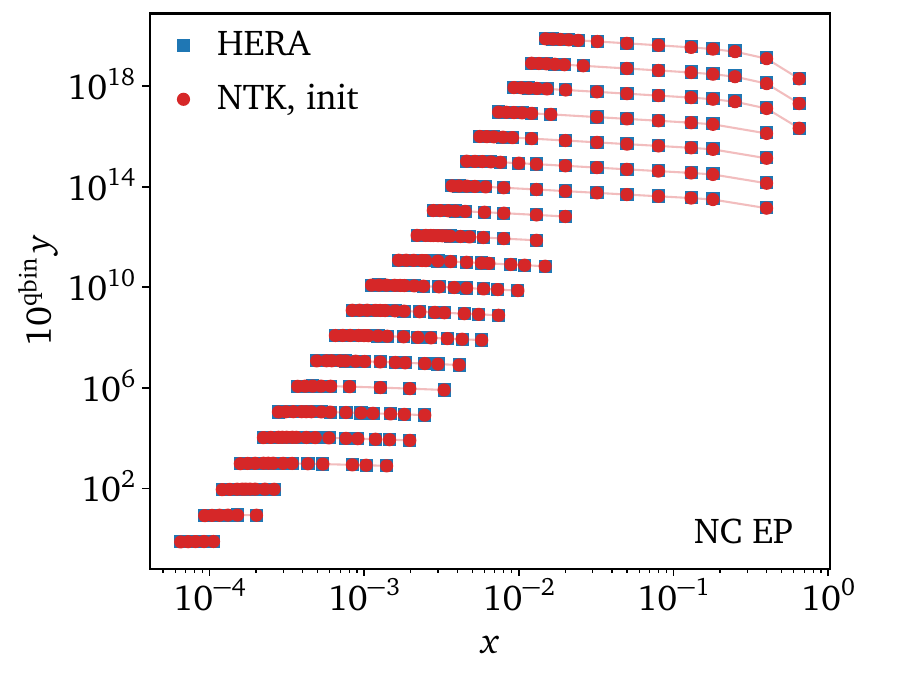}
    \includegraphics[width=0.49\textwidth]{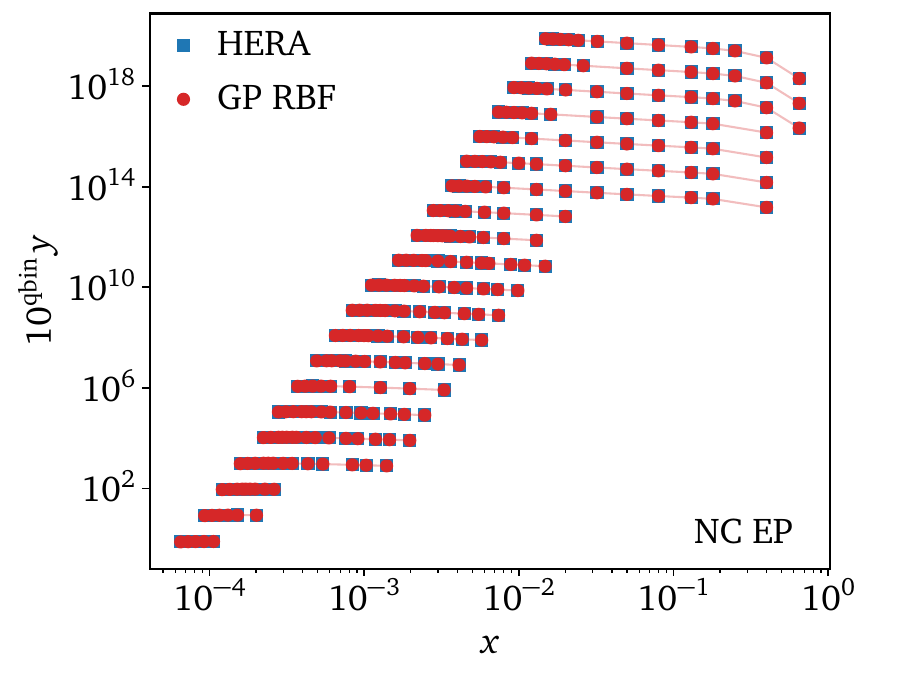}
    \caption{Comparison between the original HERA theory prediction and modified predictions in which 
    the $T_3$ contribution is replaced by the BCDMS-trained model reconstruction. The upper panels show the NC EM dataset, and the lower panels the NC EP dataset. The left panels use the NTK evaluated at initialisation for reconstruction, while the right panels use the RBF GP reconstruction. All results refer to the same HERA Level-1 pseudo-data, generated from the closure-test truth, used in Tab.~\ref{tab:hera_f3_nc_em_ep}.}
    \label{fig:mod_comp_q2}
\end{figure}
Figure~\ref{fig:mod_comp_q2} shows the same comparison at the level of the full observable-space predictions.
The original HERA NC theory prediction is compared with modified predictions in which the $T_3$ contribution is 
replaced by the one obtained from the BCDMS-trained model.
The modified predictions remain close to the native HERA prediction over the displayed kinematic range 
for both the NC EM and NC EP datasets, including regions at smaller $x$ and larger $Q^2$ than those directly 
covered by the BCDMS training data.

\begin{figure}[t!]
    \centering
    \includegraphics[width=0.45\textwidth, page=1]{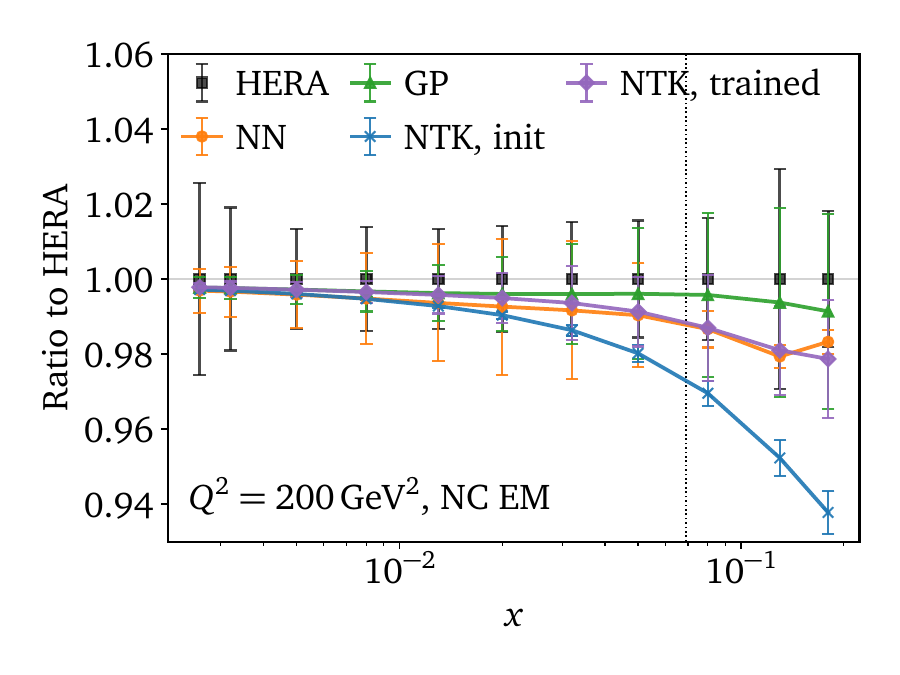}
    \includegraphics[width=0.45\textwidth, page=2]{figs/hera/NC_EM_EP_all_methods_ratio_native.pdf}
    \caption{Ratio of the modified HERA prediction to the native HERA theory prediction for the NC EM dataset 
    at fixed $Q^2$. The modified prediction is obtained by replacing the native $T_3$ contribution with the 
    BCDMS-trained model prediction. The black points correspond to the native prediction normalised to itself, 
    with experimental HERA uncertainties divided by the native theory prediction. The other points show the 
    modified-to-native ratio, with the model-spread uncertainty propagated through the HERA FK map and 
    normalised by the same native prediction. The orange points use the $\chi^2$-trained NN reconstruction, 
    while the green triangles show the RBF GP reconstruction, the blue crosses show the results obtained from 
    an NTK used at initialisation and the purple diamonds show the results obtained from an NTK after training. 
    The vertical dashed line at $x\simeq0.07$ marks the lower edge of the BCDMS-covered region. All results 
    refer to the same HERA Level-1 pseudo-data, generated from the closure-test truth, used in 
    Tab.~\ref{tab:hera_f3_nc_em_ep}.}
    \label{fig:ratio_nc_em}
\end{figure}

To examine the comparison more locally, Fig.~\ref{fig:ratio_nc_em} shows fixed-$Q^2$ slices of the 
modified-to-native ratio,
\begin{align}
\frac{T_{\rm pred}}{T_{\rm HERA}} \eqcomma
\end{align}
where $T_{\rm pred}$ is obtained by replacing only the $T_3$ contribution and $T_{\rm HERA}$ denotes the native 
HERA theory prediction.
The black squares, therefore, lie at unity by construction, with $1\sigma$ intervals corresponding to the 
experimental HERA uncertainty normalised by the native prediction.
The coloured points show the effect of the $T_3$ replacement for the different models, with uncertainties given 
by the model-spread contribution propagated to HERA observable space.
The slice at $Q^2=200\,{\rm GeV}^2$ lies within the approximate BCDMS-covered range, while 
the $Q^2=500\,{\rm GeV}^2$ slice probes a more extrapolative regime.
The NN, GP and trained-NTK substitutions remain compatible with the original HERA prediction within 
the displayed uncertainties,
while the NTK at initialisation deviates from it by several percent, beyond the experimental uncertainties, 
already for $x\gtrsim0.07$, consistently with its poor reconstruction in the data region.
Tests performed on the NC EP dataset yielded results similar to those of the NC EM dataset.
Thus, we do not show them here.
This indicates that within this closure-test setup, the BCDMS-trained reconstructions capture the dominant 
$T_3$ structure relevant to these HERA observables.
A full phenomenological validation would require a simultaneous global fit in which all relevant PDF combinations 
are allowed to vary, and their correlations are propagated consistently. 
The present test is more limited, in that it checks only whether the isolated $T_3$ replacement produces a 
visible inconsistency when inserted into related HERA DIS observables.

\section{Conclusion and Outlook}
\label{sec:summary}
In this work, we have compared the uncertainty estimates of Gaussian processes, neural-network ensembles and empirical neural tangent kernel approximations in the simplest instance of a PDF determination: the reconstruction of the non-singlet triplet distribution, denoted as $T_3(x,Q_0^2)$, from DIS structure-function data, related to the PDF by an ill-conditioned linear forward map. 
Working in a closure-test setting, with correlated experimental uncertainties, allowed us to compare each reconstruction and its uncertainty directly with the known underlying truth. 
This provides an intermediate benchmark between direct toy-function regression and a global PDF determination: the underlying PDF is known, but it is constrained only indirectly through observable-space data.

Our central result is that reconstruction accuracy, uncertainty magnitude, and uncertainty calibration are distinct properties. 
Several of the methods reproduce the main structure of the underlying distribution in the data-supported region, while assigning substantially different uncertainty bands and covariance structures to their predictions. 
Conversely, a broad pointwise band does not guarantee that the predictive covariance aligns with the observed residuals. 
Central-value agreement and visual inspection of uncertainty bands are therefore insufficient for validating uncertainty estimates in this inverse problem.

The neural-network comparisons demonstrate that the loss function affects the uncertainty estimate more strongly than the central reconstruction. 
On the fixed $L_1$ realisation, the MSE, heteroscedastic-MSE, and covariance-$\chi^2$ fits give similar central predictions in observable space, but their PDF-space uncertainties differ markedly. 
The standard MSE does not use the experimental covariance and produces limited ensemble spread, especially towards small-$x$ extrapolation. 
The heteroscedastic likelihood learns a much broader pointwise variance, but its diagonal-covariance assumption cannot capture correlated experimental uncertainties. 
Its large uncertainty band should therefore not be interpreted as a calibrated PDF uncertainty. 
The $\chi^2$ ensemble produces the strongest growth of uncertainty towards small $x$. 
On noise-free $L_0$ data, the three reconstructions become nearly indistinguishable, but they share a common small-$x$ deviation from the truth that is not covered by their uncertainty bands. 
Since the data are noise-free, this deviation is due to the small-$x$ preprocessing: $\alpha\simeq1$ forces $xT_3\propto x$ as $x\to0$. 
Removing the data noise therefore reduces the dependence on the loss, but not the dependence of the extrapolation on the prior.

The endpoint prefactor has a particularly important role. 
For the NNs, the raw and preprocessed fits are similar where data provide direct constraints, but differ qualitatively outside that region. 
Without preprocessing, the raw ensembles develop very large small-$x$ uncertainties and unstable large-$x$ behaviour. 
The prefactor suppresses these variations and enforces the expected endpoint limits. This improves stability, but it also means that the resulting extrapolation and its uncertainty are partly determined by the assumed functional form rather than inferred from the data. 
This is made explicit by varying the small-$x$ exponent at fixed $\beta$: since the trained exponents remain close to their initial values, 
fits with $\alpha=0.5$ and $\alpha=1.0$ agree in the data region but differ substantially below $x\simeq0.07$, both in their central values and in their uncertainties, 
and this dependence persists for fits to $L_2$ replicas. 
The small-$x$ uncertainty bands are therefore conditional on the preprocessing.

The same effect is apparent, and even more pronounced, for the Gaussian processes. 
Without endpoint preprocessing, the different kernels give visibly different central predictions and uncertainties, and none reconstructs the peak uniformly well. 
Including the endpoint prefactor substantially improves the central reconstruction, with the preprocessed RBF GP providing one of the closest central predictions among the representative models. 
At the same time, the GP uncertainties vanish at the endpoints by construction, since the prior standard deviation of $xT_3$ is proportional to $x^{\alpha}(1-x)^{\beta}$, 
so that at small $x$ they do not cover the truth.
These narrow bands should therefore be interpreted as prior-conditioned extrapolations, not as independently validated precision. 
The GP results illustrate that a physically motivated prior can improve reconstruction while simultaneously making the reported extrapolation uncertainty more dependent on modelling assumptions.

The empirical NTK provides a complementary example. 
Using the NTK of a single network as the covariance of a GP prior, with the network output as prior mean, and conditioning it on the data 
yields an uncertainty estimate that inherits the features learned by the network. 
When the kernel is evaluated at random initialisation, the reconstruction substantially underestimates the peak and its uncertainty is far too small ($R_b\simeq9.5$), 
so that the initialisation NTK is not a faithful surrogate for the trained nonlinear network. 
Most of the improvement in the central prediction is already visible when the NTK is evaluated after 100 training epochs, showing that the relevant local geometry changes early in optimisation. 
Extending training to 5000 epochs changes the central reconstruction only modestly, but produces a narrower extrapolation uncertainty 
and an uncertainty in the interpolation region that is compatible with, or slightly larger than, the observed deviation from the truth. There are natural extensions for these NTK uncertainties: the variation between over- and under-estimating uncertainty throughout training deserves further study, in particular whether it allows for a controlled calibration by appropriate termination of training. In addition, since a single network is used, this uncertainty does not include the variability of the empirical NTK across random initialisations.

The normalised bias $R_b$ provides a quantitative test of these uncertainties in the interpolation region, whose interpretation follows from the cumulative structure of the closure-test levels. 
The $L_1$ ensemble, trained on a single data realisation, does not include the data uncertainty, and accordingly gives $R_b\simeq2.8$. 
All setups that include it, either through $L_2$ replicas or through the experimental covariance in the likelihood, as for the GPs and the trained NTK, 
give values compatible with a faithful uncertainty at the 95\% confidence level, or, for the Mat\'ern kernel with $\nu=0.5$ and the trained NTK, marginally below this range, 
indicating slightly conservative uncertainties. 
The only setup that includes the data uncertainty and nevertheless underestimates it is the NTK at initialisation. 
Replacing the full ensemble covariance by its diagonal changes $R_b$ substantially and in a model-dependent direction, showing that calibration depends not only on marginal band widths, 
but also on how the deviations project onto the correlated eigenmodes of the predictive covariance. 
However, since the ensemble covariances are dominated by a small number of eigenvectors, $n_{\rm eig}=4$--$19$, $R_b$ probes only a few directions in data space. 
On a single realisation, where its statistical precision is about $1/\sqrt{2n_{\rm eig}}$, it can therefore exclude a substantial underestimate of the uncertainty, but cannot establish its faithfulness.
Repeating the test on $N_{L_1}$ independent $L_1$ realisations, i.e.\ on different ``runs of the universe'', would reduce the statistical uncertainty on the average $R_b$ as $1/\sqrt{N_{L_1}}$.
Moreover, being evaluated only in the interpolation region, $R_b$ does not test the extrapolation.

Finally, we tested the extrapolation of the reconstructed $T_3$ by propagating it to two HERA neutral-current reduced-cross-section datasets, 
which extend to values of $x$ below the BCDMS range, while keeping all other PDF contributions fixed. 
Since the HERA Level-1 pseudo-data are generated from the same truth used for the baseline prediction, any deviation of the reconstruction from the truth is expected to increase the $\chi^2$, although individual realisations can give negative shifts. 
For the NC EM dataset, with the experimental covariance, the shifts are positive or compatible with zero and remain below one standard deviation for all reconstructions 
except the NTK at initialisation, which gives a shift of about five standard deviations; for the NC EP dataset all shifts are negligible.
The extrapolation test therefore singles out the same setup that fails the interpolation diagnostic. 
Its discriminating power is nevertheless limited by the small contribution of $T_3$ to the complete HERA observable and by the sizeable experimental uncertainties in parts of the kinematic range, so that the result should be regarded as a consistency check of the reconstructed extrapolation rather than as a full validation of its uncertainty.

The two tests allow us to summarise which models perform well in each region.
In the data (interpolation) region, the GPs with the Gibbs, RBF and Mat\'ern ($\nu=1.5$ and $2.5$) kernels and the NN ensemble trained on $L_2$ replicas
give uncertainties compatible with a faithful one, while the Mat\'ern kernel with $\nu=0.5$ and the trained NTK give slightly conservative uncertainties;
among these, the preprocessed RBF GP provides the closest central reconstruction.
The NN ensemble trained on a single $L_1$ realisation, which by construction lacks the data uncertainty, and the NTK at initialisation underestimate the uncertainty.
In the extrapolation region, all reconstructions except the NTK at initialisation pass the HERA test, but their uncertainties behave very differently.
The $\chi^2$-trained NN ensemble fitted to $L_2$ replicas is the setup whose uncertainty grows most below $x\simeq0.07$ and covers the truth,
the trained NTK gives a narrower band, and the MSE-trained ensembles remain narrow,
while the uncertainties of the preprocessed GPs vanish as $x\to0$ and do not cover the truth.
In all cases the small-$x$ extrapolation is conditional on the assumed preprocessing exponent.

Taken together, our results identify the ingredients that determine whether ML-based uncertainty estimates are faithful in this inverse problem, 
independently of the quality of the central reconstruction. 
First, the data uncertainty must be propagated, either through Monte Carlo replicas or through a likelihood that includes the full experimental covariance: 
an ensemble trained on a single data realisation does not include it, and a loss that learns only pointwise variances cannot account for correlated experimental uncertainties. 
Second, kernel approximations of a trained network must be evaluated at the trained parameters, since the NTK at initialisation is not a faithful surrogate. 
Third, in the extrapolation region the uncertainty is largely determined by the functional prior, encoded in the kernel or in the endpoint preprocessing, 
and should be regarded as conditional on it. 
Closure tests, in which the underlying truth is known, combined with covariance-aware diagnostics and tests outside the data-supported region, 
are therefore essential to validate uncertainties in ML-based approaches to inverse problems.

For PDF determinations, we expect these conclusions to generalise to global fits.
Single-flavour determinations of the non-singlet triplet have historically been a reliable proxy for the full problem, on which new methodologies were first validated
before being extended to global analyses~\cite{DelDebbio:2007ee,Ball:2008by,Candido:2024hjt,Chiefa:2025cap},
and the features that drive our results, namely an ill-conditioned forward map, correlated experimental uncertainties, and a limited coverage in $x$ that requires extrapolation,
are common to all PDF determinations.
For inverse problems of a different nature, in which another underlying law is inferred from data, we cannot conclude that the same models would perform in the same way,
since the outcome depends on the forward map, on the structure of the data uncertainties and on the available prior information.
What does generalise is the procedure: a closure test with a known underlying law and realistic correlated uncertainties, the comparison of the $L_0$, $L_1$ and $L_2$ levels,
a covariance-aware calibration diagnostic in the data region, and a test on data lying outside the region used for training.
This procedure can be adopted in any other study in which ML methods are used to solve an inverse problem.

A necessary next step is to repeat the analysis over many independent $L_1$ closure-test realisations. 
This would permit direct measurements of pointwise and simultaneous coverage, calibration along covariance eigenmodes, and the distribution of $R_b$, rather than relying on one pseudo-dataset, 
and, by averaging over realisations as in the bias--variance ratio of Refs.~\cite{DelDebbio:2021whr,NNPDF:2021njg}, would compensate for the small number of directions probed by $R_b$ on a single realisation,
reducing its statistical uncertainty as $1/\sqrt{N_{L_1}}$.
The $L_2$ analysis should also be extended towards a closer analogue of the complete NNPDF methodology, including systematic studies of replica number, preprocessing variation, in particular of the small-$x$ exponent, validation and stopping prescriptions, and training stochasticity. 
For GPs, kernel and preprocessing dependence should be incorporated into the uncertainty assessment, for example through hierarchical inference or model averaging.
For NTKs, a quantitative study of the kernel evolution during training for the present architecture, together with the inclusion of the variability 
of the empirical NTK across random initialisations, 
would clarify which uncertainty information is retained by the tangent approximation. 
A further extension is to repeat the study with built-in inconsistent data, as done in Ref.~\cite{Barontini:2025lnl}.
There it was shown that an NN ensemble trained on moderately inconsistent data corrects for the inconsistency;
it would be interesting to establish whether this also holds for the other models considered here, in particular for GPs and NTK-based approximations,
whose uncertainty is determined by the likelihood and the kernel rather than by an ensemble of fits.
Finally, extending the benchmark to simultaneous multi-flavour fits, with experimental and theory covariances propagated across several observables, will determine which of these behaviours persist in realistic global PDF determinations.

\section*{Acknowledgments}
We are very grateful to Amendeo Chiefa and Luigi Del Debbio for helpful comments on the draft. 
MU is supported by the European Research Council under the European Union’s Horizon
2020 research and innovation Programme (PBSP, Grant agreement n.950246). SK and MU are partially
supported by the STFC consolidated grants ST/X000664/1 and UKRI4224.
NE is funded by the Infosys-Cambridge AI Centre.
\newpage

\appendix
\section{Hyperparameters}
\label{app:hyper}
This appendix contains tables for the different hyperparameters used in the analysis.
All NN and NTK training runs use a fixed training length of 5000 epochs. No early-stopping criterion 
is applied in this controlled comparison.
Tab.~\ref{tab:hyperparameter_nn} contains the hyperparameters used for the NN setup, NTK initialisation 
and NTK after training.
We use $\log x$ as the sole NN input, rather than the pair $(x,\log x)$ used by NNPDF, to keep the 
parametrisation simple for this reduced problem, which involves a single PDF combination and a more 
restricted dataset than a global fit.
Tab.~\ref{tab:hyperparameter_gp} shows the hyperparameters used for GP with different kernels.
\begin{table}[htpb]
    \setlength{\tabcolsep}{10pt}
    \centering
    \begin{small} \begin{tabular}{l|c}
        \toprule
         Hyperparameter & value\\
         \midrule
         epochs & 5000 \\
         optimizer & Adam\\
         learning rate & $5 \cdot 10^{-4}$\\
         dropout & 0.001\\
         \midrule
         number of layers & 3 \\
         hidden features & 32 \\
         activation function & GELU\\
         ensemble members & 100 \\
         \bottomrule
    \end{tabular} \end{small}
    \caption{Hyperparameters for the NN architecture, using either a $\chi^2$, MSE, or heteroscedastic MSE loss.}
    \label{tab:hyperparameter_nn}
\end{table}
\begin{table}[htpb]
    \setlength{\tabcolsep}{10pt}
    \centering
    \begin{small} \begin{tabular}{l|c}
        \toprule
         Hyperparameter & value\\
         \midrule
         $\alpha$ & uniform, $[0.1, 1.1]$ \\
         $\beta$ & uniform, $[1.7, 3.5]$ \\
         $l_0$ & uniform, $[10^{-6}, 10]$ \\
         $\sigma$ & uniform, $[10^{-6}, 10]$ \\
         $\nu$ & $\in [0.5, 1.5, 2.5]$ \\
         \midrule
         \# chains & 4\\
         tune & 500 \\
         draws & 2000 \\
         target acceptance & 0.97 \\
         \bottomrule
    \end{tabular} \end{small}
    \caption{Hyperparameter prior ranges used for the GP fits. The preprocessing exponents $\alpha,\beta$ are included only for the endpoint-preprocessed GP setup. The length-scale parameter is denoted by $\ell$ for the stationary kernels and by $\ell_0$ for the Gibbs kernel.}
    \label{tab:hyperparameter_gp}
\end{table}

\section{Endpoint-prefactor sensitivity}
\label{app:scan}
\begin{figure}[t!]
    \centering
    \includegraphics[width=0.32\textwidth, page=2]{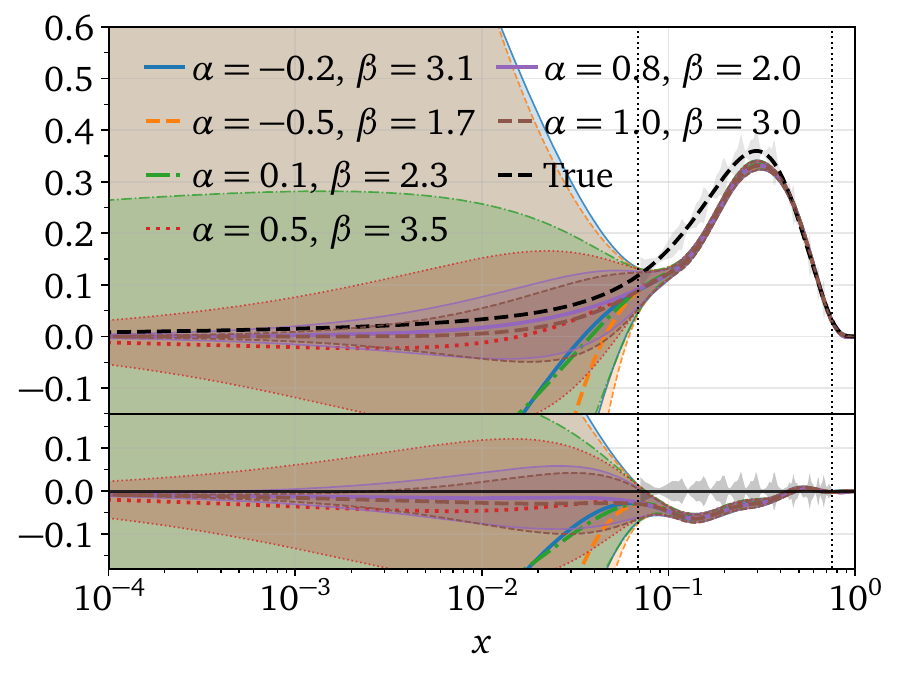}
    \includegraphics[width=0.32\textwidth, page=2]{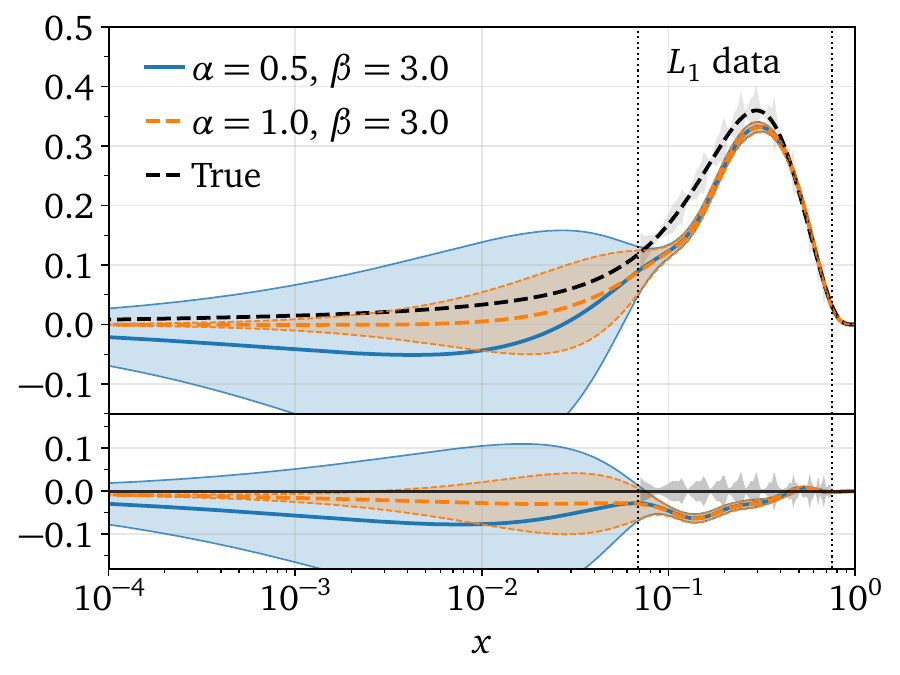}
    \includegraphics[width=0.32\textwidth, page=2]{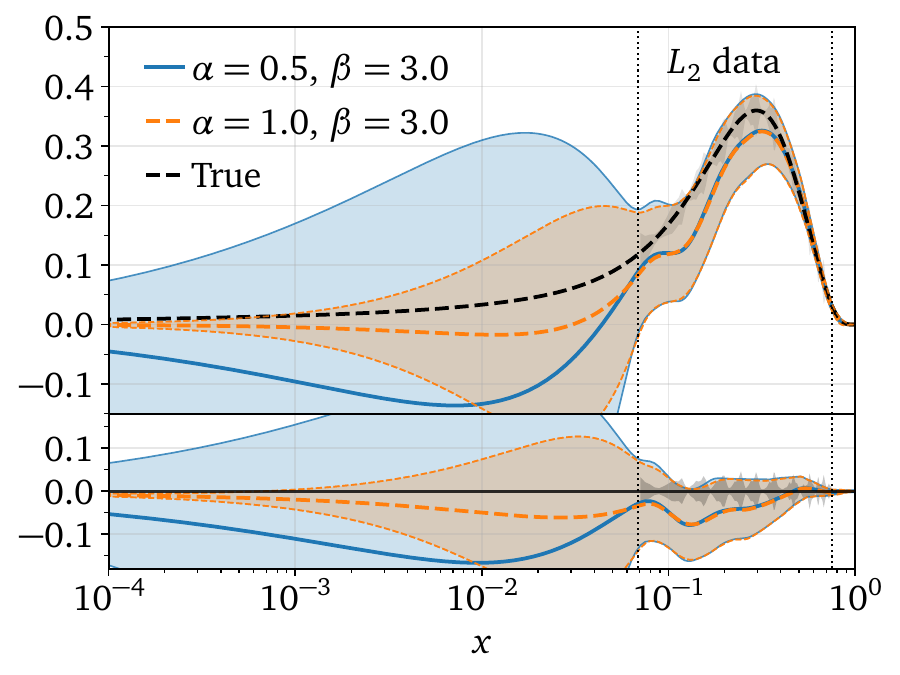}
    \caption{Sensitivity of the $\chi^2$-trained NN reconstruction to the endpoint exponents. Left: fixed values of $\alpha$ and $\beta$ for $L_1$ data. Centre and right: controlled comparison of fixed $\alpha=0.5$ and $\alpha=1.0$ at fixed $\beta=3.0$, for $L_1$ (centre) and $L_2$ (right) ensembles, with all other settings identical within each pair. Lower panels show residuals to the closure-test truth.}
    \label{fig:alpha_scan_fixed_beta}
\end{figure}

In this appendix we study the sensitivity of the results to the endpoint prefactor of Eq.~\eqref{eq:scaling}. 
Fig.~\ref{fig:alpha_scan_fixed_beta} (left) shows the $\chi^2$-trained NN fitted to $L_1$ data for several 
fixed values of $\alpha$ and $\beta$, including values outside the default trainable range. 
The interpolation region is essentially unaffected, while the transition into the extrapolation region and the 
extrapolation itself depend strongly on the exponents. 
We find the same behaviour for fits to $L_2$ replicas and when the exponents are trainable but started from 
the same set of initial values, since the trained exponents remain close to their initial values.

To isolate the small-$x$ dependence, the central and right panels of Fig.~\ref{fig:alpha_scan_fixed_beta} 
compare $\chi^2$-trained NN ensembles with fixed $\alpha=0.5$ and $\alpha=1.0$ at fixed $\beta=3.0$, 
for $L_1$ and $L_2$ data respectively, with all other settings identical within each pair. 
Since the prefactor multiplies $xT_3$, these choices correspond to Regge-like and flat reference powers for $T_3$. 
The reconstructions are similar in the data-supported region, but their central values and uncertainty bands differ 
substantially below $x\simeq0.07$, 
and this dependence persists in the $L_2$ setup: data-replica variation does not remove the preprocessing dependence 
in the extrapolation region.

\begin{figure}[b!]
    \centering
    \includegraphics[width=0.32\textwidth, page=3]{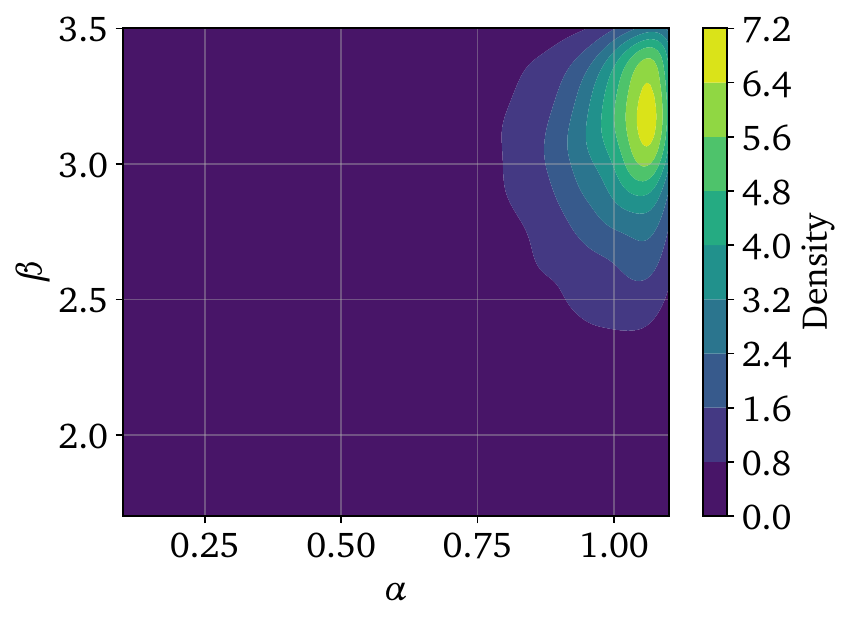}
    \includegraphics[width=0.32\textwidth, page=3]{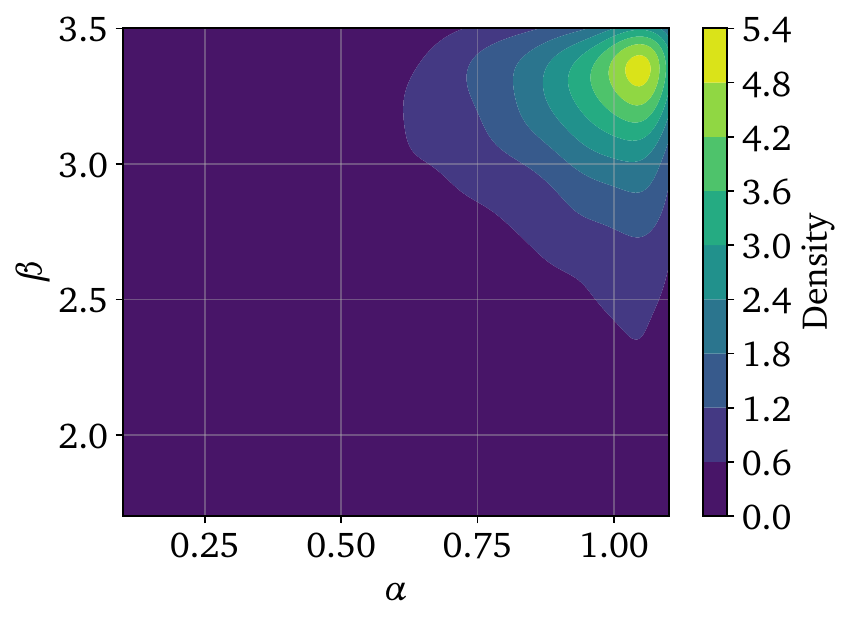}
    \includegraphics[width=0.32\textwidth, page=3]{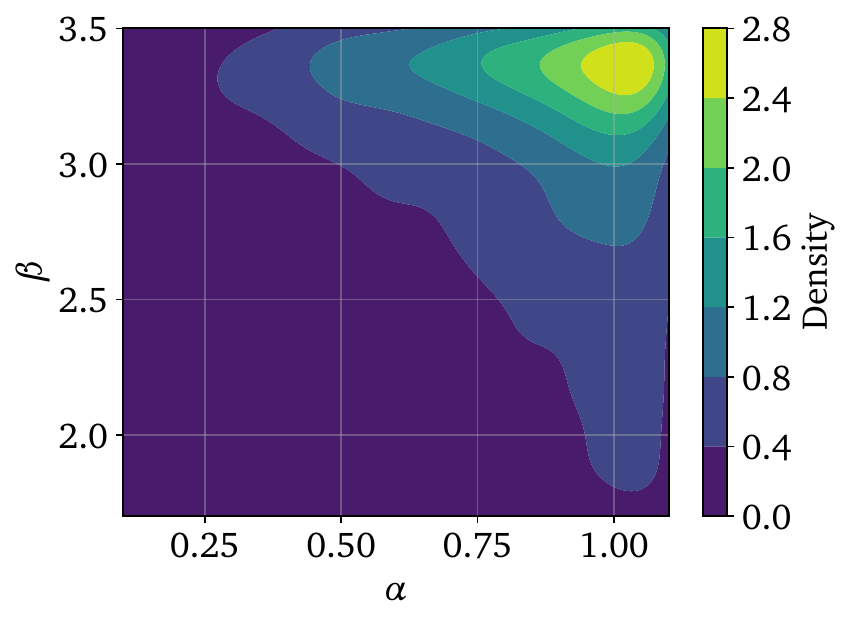}
    \caption{Joint posterior distributions of the endpoint exponents $\alpha$ and $\beta$ in the endpoint-preprocessed GP fits with the Mat\'ern ($\nu=0.5$, left), Gibbs (centre) and RBF (right) kernels.}
    \label{fig:gp_posteriors}
\end{figure}

Finally, Fig.~\ref{fig:gp_posteriors} shows the joint posterior distributions of the endpoint exponents 
in the endpoint-preprocessed GP fits. 
For all kernels the posterior peaks at $\alpha\simeq1$, close to the upper edge of the prior range, and 
at $\beta\simeq3.1$ for the Mat\'ern kernel with $\nu=0.5$ 
and $\beta\simeq3.3$--$3.4$ for the other kernels; the Mat\'ern kernels with $\nu=1.5$ and $2.5$, not shown, 
interpolate between the Mat\'ern $\nu=0.5$ and the RBF results. 
We therefore adopt $\alpha=1.0$ and $\beta=3.0$ as starting values in the NN and NTK setups, although this 
choice remains a modelling assumption. 
The posteriors become broader as the kernel becomes smoother, confirming that the endpoint prefactor and 
the covariance kernel should be interpreted jointly rather than as independent modelling choices.

\section{Supplementary material}
This appendix provides supplementary comparisons related to the simplified NNPDF-style replica 
setup used in the main text and investigates the distribution of the used $L_1$ data generation 
by calculating the pull for $1000$ independent $L_1$ random samples.

\subsection*{$L_2$ data comparisons}
\label{app:L2}
\begin{figure}[t!]
    \centering
    \includegraphics[width=0.49\textwidth, page=1]{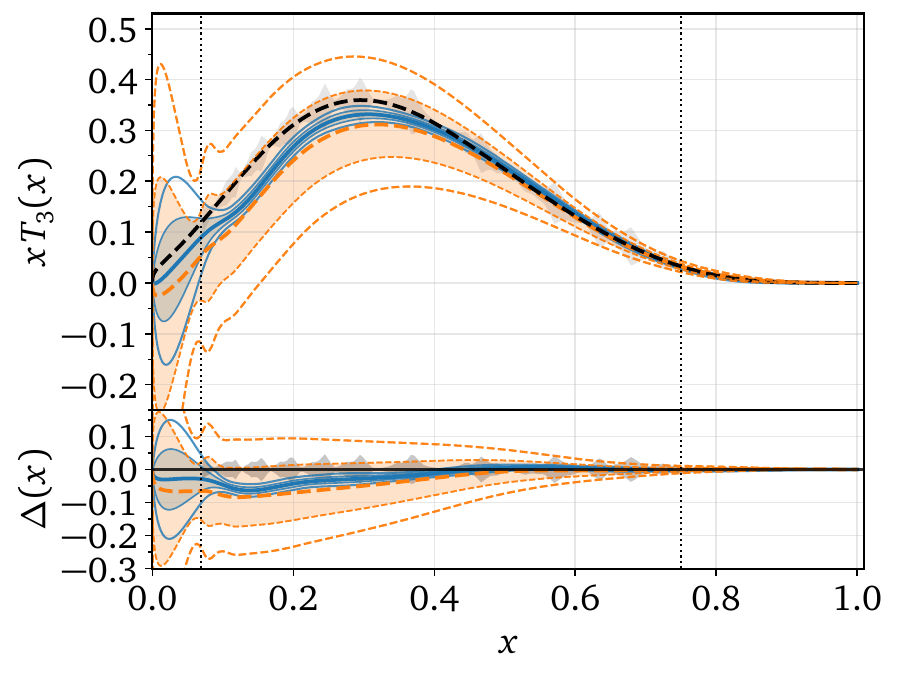}
    \includegraphics[width=0.49\textwidth, page=2]{figs/app/L2_NN2.pdf}
    \caption{Supplementary comparison of a fixed-$L_1$ NN $\chi^2$ fit with an $L_2$-replica $\chi^2$ fit. The $L_2$ setup follows the simplified NNPDF-style replica construction described in Sec.~\ref{sec:t3model}. Results are shown on a linear $x$ scale (left) and logarithmic $x$ scale (right), with residuals to the closure-test truth in the lower panels. The shaded area indicates the $68\%$ confidence intervals, while the lines bound the $95\%$ confidence intervals.}
    \label{fig:L2_NN2}
\end{figure} 
\begin{figure}[b!]
    \centering
    \includegraphics[width=0.49\textwidth, page=1]{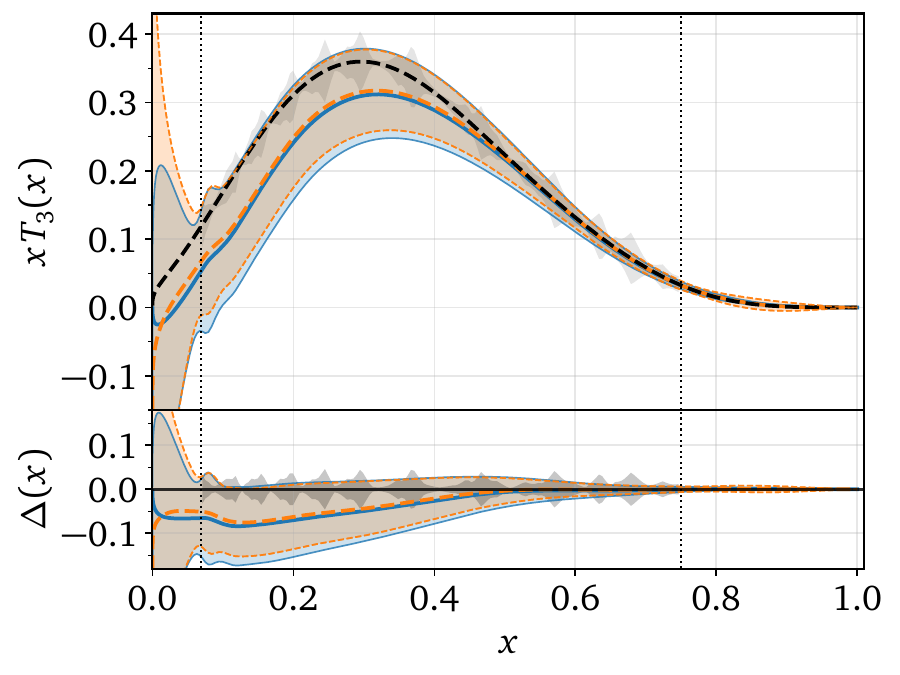}
    \includegraphics[width=0.49\textwidth, page=2]{figs/app/L2_scale_init.pdf}
    \caption{Supplementary comparison including an alternative, more NNPDF-like $L_2$ setup. In the $\chi^2$ $L_2$ varied-prefactor fit, the endpoint preprocessing exponents are initialised independently for each replica. The comparison illustrates the effect of adding replica-dependent preprocessing variation.}
    \label{fig:L2_init}
\end{figure} 
Figure~\ref{fig:L2_NN2} shows how the NN $\chi^2$ results change when moving from 
fixed-$L_1$ training to $L_2$-replica training, including the $68\%$ (shaded area) and 
the $95\%$ (bounded by lines) confidence intervals.
Figure~\ref{fig:L2_init} then compares a more NNPDF-like variant, in which additional replica-dependent variation 
is introduced through the endpoint preprocessing (label: $\chi^2$, $L_2$, pref),
to the more naive $L_2$ implementation ($\chi^2$, $L_2$), which uses a fixed preprocessing across all replicas.
These comparisons are not used as separate calibration diagnostics but serve as robustness checks for the central 
reconstruction and the ensemble spread.
\subsection*{Distribution of $L_1$ data}
\begin{figure}[t!]
    \centering
    \includegraphics[width=0.49\textwidth, page=1]{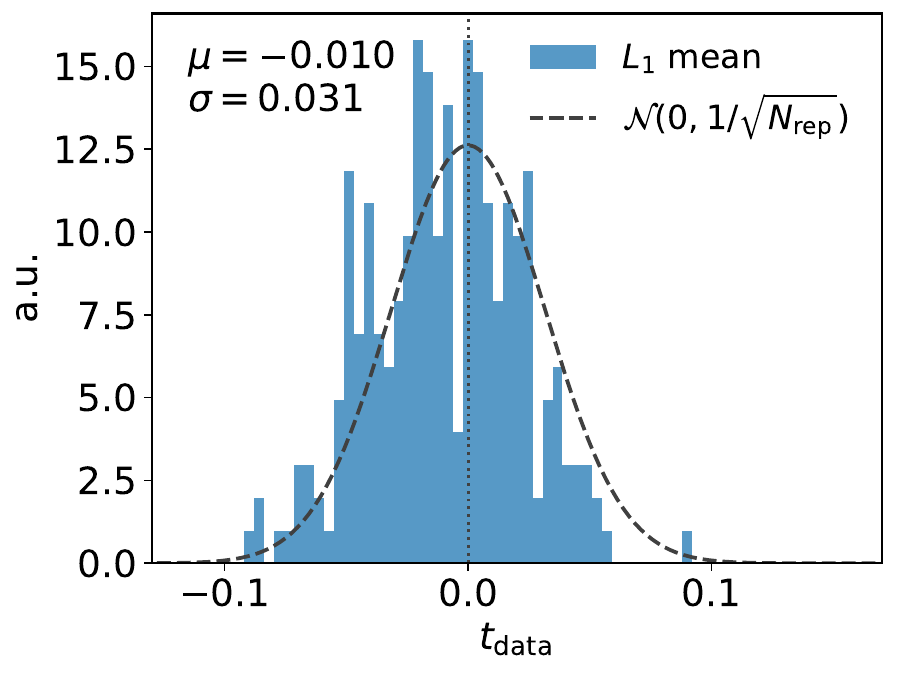}
    \caption{Pull distribution for $1000$ independently generated $L_1$ data samples.}
    \label{fig:data_pull}
\end{figure} 
Regarding the issue with the underestimated predictions when using one $L_1$ replica sample as seen 
in Fig.~\ref{fig:NN_loss}, we investigated this behaviour further by generating $1000$ independent $L_1$ samples.
To check their calibration with respect to the underlying truth $y$ and the experimental covariance 
matrix $C_{\rm exp}$ we show the corresponding pull distribution defined as
\begin{align}
    t_{\rm data} = \frac{\langle y_{L_1} \rangle - y_{\rm theory}}{\sqrt{C_{\mathrm{exp}, ii}}} \eqperiod
\end{align}
\begin{figure}[b!]
    \centering
    \includegraphics[width=0.49\textwidth, page=1]{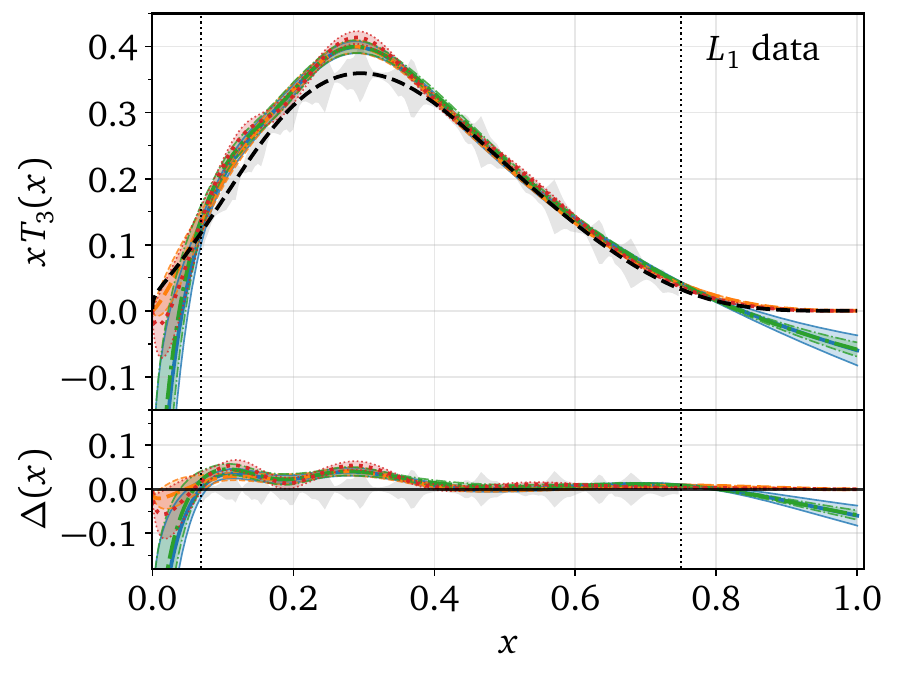}
    \includegraphics[width=0.49\textwidth, page=2]{figs/app/NN_L1_arc_newseed.pdf}
    \caption{Comparison of $L_1$ data using a NN with MSE or $\chi^2$ loss, including endpoint scaling (full) or not including it (raw). Compared to Fig.~\ref{fig:NN_L1_arc}, the raw results depend heavily on the random seed in the $L_1$ data generation, while the full setup can recover the underlying truth equally well.}
    \label{fig:NN_diffseed}
\end{figure} 
When using $N_{\rm rep} = 1000$ we expect the distribution to follow a Gaussian distribution with 
$\mathcal{N}(0, \sigma=1/\sqrt{N_{\rm rep}})$
This translates to $\sigma = 0.0316$. 
The pull distribution is shown in Fig.~\ref{fig:data_pull}.

Furthermore, to show the impact of randomisation on $L_1$ replicas, we reproduce Fig.~\ref{fig:NN_L1_arc} 
with a different random seed in the $L_1$ generation. 
The result is shown in Fig.~\ref{fig:NN_diffseed}.
We can see that the systematic underestimation of the true is indeed related to the random seed picked for $L_1$.
Additionally, the random seed, without any additional information introduced by endpoint preprocessing, has 
a significant impact on the extrapolation behaviour.

\begin{figure}[t!]
    \centering
    \includegraphics[width=0.49\textwidth, page=1]{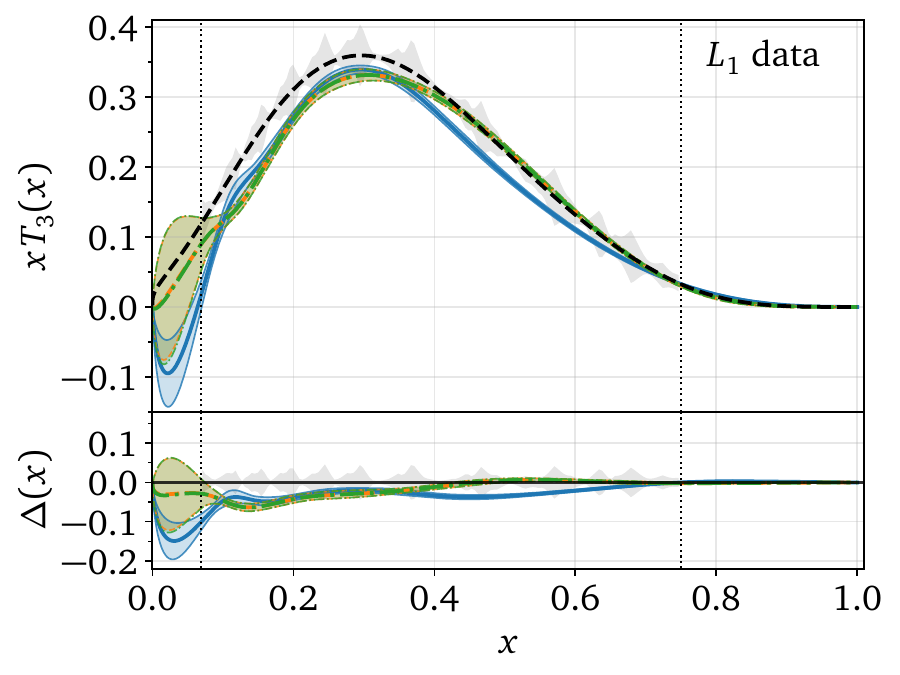}
    \includegraphics[width=0.49\textwidth, page=2]{figs/app/C_T0_test.pdf}
    \caption{Comparison of $L_1$ data using a NN with $\chi^2$ loss, including either $C_{\rm exp}$, $C_{T_0}$ or an iterated $C_{T_0}$ matrix in the loss function.}
    \label{fig:NN_ct0}
\end{figure}

Fig.~\ref{fig:NN_ct0} shows a comparison of using the $C_{\rm exp}$, $C_{T_0}$ or an iterated version of 
$C_{T_0}$ in the loss function for a NN with a $\chi^2$ loss. Using $C_{T_0}$ improves the central prediction, but iterating it adds no changes or improvements.

\bibliography{references}
\end{document}